\documentstyle[onecolumn,subfigure]{mn}
\begin{document}
\newcommand{\mincir}{\raise
-2.truept\hbox{\rlap{\hbox{$\sim$}}\raise5.truept
\hbox{$<$}\ }}
\newcommand{\magcir}{\raise
-2.truept\hbox{\rlap{\hbox{$\sim$}}\raise5.truept
\hbox{$>$}\ }}
\newcommand{\minmag}{\raise-2.truept\hbox{\rlap{\hbox{$<$}}\raise
6.truept\hbox
{$>$}\ }}
\newcommand{\be}{\begin{equation}}
\newcommand{\ee}{\end{equation}}
\newcommand{\ba}{\begin{eqnarray}}
\newcommand{\ea}{\end{eqnarray}}
\newcommand{\brr}{\begin{array}}
\newcommand{\nn}{\nonumber \\} 
\newcommand{\err}{\end{array}}
\newcommand{\bc}{\begin{center}}
\newcommand{\ec}{\end{center}}
\newcommand{\br}{\mbox{\bf r}}
\newcommand{\bv}{\mbox{\bf v}}
\newcommand{\bs}{\mbox{\bf s}}
\newcommand{\bq}{\mbox{\bf q}}
\newcommand{\bx}{\mbox{\bf x}}
\newcommand{\by}{\mbox{\bf y}}
\newcommand{\bk}{\mbox{\bf k}}
\newcommand{\tR}{\mbox{\tiny R}}
\newcommand{\tM}{\mbox{\tiny M}}
\newcommand{\tN}{\mbox{\tiny N}}
\newcommand{\tL}{\mbox{\tiny L}}
\newcommand{\lb}{{\left<\right.}}
\newcommand{\rb}{{\left.\right>}}
\newcommand{\hm}{\,h^{-1}{\rm Mpc}}
\newcommand{\vel}{\,{\rm km\,s^{-1}}}

\renewcommand{\baselinestretch}{1.0}
\input{psfig.sty} 
\title[Measuring and Modelling the Redshift Evolution
of Clustering: the HDF North]
{Measuring and Modelling the Redshift Evolution
of Clustering:  the Hubble Deep Field North}
\author[Arnouts et al.]
{S. Arnouts$^{1,2}$, S. Cristiani$^{1,3}$, L. Moscardini$^{1}$,
S. Matarrese$^{4}$, F. Lucchin$^{1}$, 
\newauthor A. Fontana$^5$ and E. Giallongo$^5$ \\ 
$^1$ Dipartimento di Astronomia, Universit\`a di Padova, vicolo 
dell'Osservatorio 5, I--35122 Padova, Italy \\ 
$^2$ Institut d'Astrophysique de Paris, CNRS, 98bis Bd. Arago, F-75014 Paris,
France \\
$^3$ Space Telescope-European Coordinating Facility, E.S.O., 
Karl-Schwarzschild Str. 2, D--85748 Garching, Germany \\
$^4$ Dipartimento di Fisica G. Galilei, Universit\`{a} di Padova, via 
Marzolo 8, I--35131 Padova, Italy \\ 
$^5$ Osservatorio Astronomico di Roma, via dell'Osservatorio,
I-00040 Monteporzio, Italy }

\maketitle

\begin{abstract}
The evolution of galaxy clustering from $z=0$ to $z\simeq 4.5$ is
analyzed using the angular correlation function and the photometric
redshift distribution of galaxies brighter than $I_{AB}\le 28.5$ in
the Hubble Deep Field North. The reliability of the photometric
redshift estimates is discussed on the basis of the available
spectroscopic redshifts, comparing different codes and investigating
the effects of photometric errors.  The redshift bins in which the
clustering properties are measured are then optimized to take into
account the uncertainties of the photometric redshifts.  The results
show that the comoving correlation length $r_0$ has a small decrease
in the range $0 \mincir z \mincir 1$ followed by an increase at higher
$z$.  We compare these results with the theoretical predictions of a
variety of cosmological models belonging to the general class of Cold
Dark Matter scenarios, including Einstein-de Sitter models, an open
model and a flat model with non-zero cosmological constant.  The
comparison with the expected mass clustering evolution indicates that
the observed high-redshift galaxies are biased tracers of the dark
matter with an effective bias $b$ strongly increasing with redshift.
Assuming an Einstein-de Sitter universe, we obtain $b\simeq 2.5$ at
$z\simeq 2$ and $b\simeq 5$ at $z\simeq 4$.  These results support
theoretical scenarios of biased galaxy formation in which the galaxies
observed at high redshift are preferentially located in more massive
halos. Moreover, they suggest that the usual parameterization of the
clustering evolution as $\xi(r,z) = \xi(r,0) \ (1+z)^{-(3+\epsilon)}$
is not a good description for any value of $\epsilon$.  A comparison
of the clustering amplitudes that we measured at $z\simeq3$ with those
reported by Adelberger et al. (1998) and Giavalisco et al. (1998),
based on a different selection, suggests that the clustering depends
on the abundance of the objects: more abundant objects are less
clustered, as expected in the paradigm of hierarchical galaxy
formation.  The strong clustering and high bias measured at $z\simeq
3$ are consistent with the expected density of massive haloes
predicted in the frame of the various cosmologies considered here.  At
$z\simeq 4$, the strong clustering observed in the Hubble Deep Field
requires a significant fraction of massive haloes to be already formed
by that epoch. This feature could be a discriminant test for the
cosmological parameters if confirmed by future observations.
  
\end{abstract}


\section{Introduction}
 
Clustering properties represent a fundamental clue about the formation
and evolution of galaxies.  Several large spectroscopic surveys have
measured the correlation function of galaxies in the local universe,
studying its dependence on morphological type or absolute magnitude
(Santiago \& da Costa 1990; Park et al. 1994; Loveday et al. 1995;
Benoist et al. 1996; Tucker et al. 1997).  Higher values of the
correlation length $r_0$ are observed for elliptical galaxies (or
galaxies with brighter absolute magnitude), while lower values are
obtained for late type galaxies (or galaxies with fainter absolute
magnitude).  This difference in the clustering strength suggests that
the various galaxy populations are not related in a straightforward
way to the distribution of the matter. To account for these
observations, one has to consider as a first approach that galaxies
are biased tracers of the matter distribution as $\xi_{\rm gal}(r) =
b^2({\cal M}) \xi_{\rm m}(r)$ (Kaiser 1984), where $\xi_{\rm gal}(r)$
refers to the spatial correlation function of the galaxies, $\xi_{\rm
m}(r)$ refers to the spatial correlation function of the mass and
$b({\cal M})$ represents the bias associated with different galaxy
populations. Here ${\cal M}$ describes the intrinsic properties of the
objects (like mass, luminosity, etc).

Deep spectroscopic surveys have made it possible to reach higher
redshifts and study the evolution of galaxy clustering.  For example
the Canada-France Redshift Survey (CFRS; Le F\`evre et al. 1996)
samples the universe up to $z\simeq 1$ while the K-selected galaxy
catalogue by Carlberg et al. (1997) reaches $z\simeq 1.5$. From these
data it has been possible to find a clear signal for evolution in the
clustering strength: the correlation length is three times smaller at
high redshifts ($z\simeq 1$) than its local value.  In addition,
Carlberg et al. (1997) have found segregation effects between the red
and blue samples similar to those observed locally.  A common approach
is to assume that the galaxy sample traces the underlying mass density
fluctuation [$b({\cal M},z)=1$, or at least $b({\cal M},z) =
constant$], and fit the clustering evolution of the mass with a
parametric form: $\xi(r,z) = \xi(r,0) (1+z)^{-(3+\epsilon)}$ (Peebles
1980), where $\epsilon$ describes the evolution of the mass
distribution due to the gravitational instability.  Such an assumption
makes it straightforward to discriminate between different
cosmological models.  From N-body simulations, Col\'{\i}n, Carlberg \&
Couchman (1997) found faster evolution in the Einstein-de Sitter
(hereafter EdS) universe than in an open universe with matter density
parameter $\Omega_{\rm 0m}=0.2$ ($\epsilon\simeq 0.8$ and
$\epsilon=0.2$, respectively).  Carlberg et al. (1997) obtained from
their data a small value of $\epsilon$ which would be quite difficult
to reconcile with an EdS universe, while Le F\`evre et al. (1996)
found a value $0 \le \epsilon \le 2$, still consistent with any
fashionable cosmological model.  However, using directly the galaxy
clustering evolution to derive the relevant properties of the mass is
a questionable practice, due to the bias acting as a complicating
factor.  Different samples select a mixture of galaxy masses and the
effective bias, which is expected in current hierarchical galaxy
formation theories to depend on redshift and mass [i.e. $b({\cal
M},z)$], plays a key role in the observed evolution of clustering.
Exciting progress in this field has been achieved with the recent
discovery of a large number of galaxies at $z\simeq 3$ (Lyman-Break
Galaxies, hereafter LBGs) using the U-dropout technique (Steidel et
al. 1996).  For the first time, the high-$z$ universe is probed via a
population of quite ``normal'' galaxies in contrast with the previous
surveys dominated by QSOs or radio galaxies. The LBG samples offer the
opportunity to estimate in a narrow time-scale ($2.6
\le z \le 3.4$) number densities, luminosities, colours, sizes, 
morphologies, star formation rates (SFR), chemical abundances,
dynamics and clustering of these primordial galaxies.  By using
different catalogues and statistical techniques, Giavalisco et
al. (1998, hereafter G98) and Adelberger et al. (1998, hereafter A98)
have measured the correlation length $r_0$ of this population. The
values they found are at least comparable to that of present-day
spiral galaxies ($r_0=2-4\ h^{-1}$ Mpc when an EdS universe is
assumed). Such a strong clustering at $z\simeq 3$ is inconsistent with
clustering evolution modeled in terms of the $\epsilon$ parameter for
any value of $\epsilon$ (G98). By comparing the correlation amplitudes
with the predictions for the mass correlation, G98 and A98 obtained
(for an EdS universe) a linear bias $b\simeq 4.5$ and $b\simeq 6$,
respectively. These results suggest that the LBGs formed
preferentially in massive dark matter haloes.

An alternative way to extend the present information over a larger
range of redshifts is to use the photometric measurements of redshifts
in deep multicolor surveys. This technique, based on the comparison
between theoretical (and/or observed) spectra and the observed colours
in various bands, makes it possible to derive a redshift estimate for
galaxies which are one or two magnitudes fainter than the deepest
limit for spectroscopic surveys (even with 10 m-class telescopes).

An optimal combination of deep observations and the photometric
redshift technique has been attained with the Hubble Deep Field (HDF)
North.  Photometric redshifts have been used to search for
high-redshift galaxies (Lanzetta, Yahil \& Fern\'andez-Soto 1996) and
investigate the evolution of their luminosity function and SFR
(Sawicki, Lin \& Yee 1996; Madau et al. 1996; Gwyn \& Hartwick 1996;
Franceschini et al. 1998), their morphology (Abraham et al. 1996; van
den Bergh et al. 1996; Fasano et al. 1998) and clustering properties
(Connolly, Szalay \& Brunner 1998; Miralles \& Pell\'o 1998;
Magliocchetti \& Maddox 1999; Roukema et al. 1999). A critical issue
is the statistical uncertainty of the photometric redshifts which
strongly depends on the number of bands following at the various
redshifts the main features of a galaxy spectral energy distribution
(hereafter SED), in particular the 4000 \AA\ break and the 912 \AA\
Lyman break.

The aim of this paper is to measure the galaxy clustering evolution in
the full redshift range $0\le z \le 4.5$, using the photometric
redshifts of a galaxy sample with $I_{AB} \le 28.5$ in the HDF North
(including infrared data, i.e.  Fern\'andez-Soto, Lanzetta \& Yahil
1999, hereafter FLY99) and carry out an extended comparison of the
results with the theoretical predictions of different current galaxy
formation scenarios based on variants of the Cold Dark Matter model.
This comparison will be performed using the techniques introduced by
Matarrese et al. (1997) and Moscardini et al. (1998), which allow a
detailed modelling of the evolution of galaxy clustering, accounting
both for the non-linear dynamics of the dark matter distribution and
for the redshift evolution of the galaxy-to-mass bias factor.

Our sample probes a population fainter than the spectroscopic LBGs and
an inter-comparison of their clustering properties will be useful to
address the differences in the nature of the two populations.
However, the photometric redshift approach should be used with some
caution when reaching such faint limits. In fact, uncertainties and
systematic errors are expected to be larger than those estimated in
the comparison of photometric and spectroscopic redshifts, which is
typically limited to $I_{AB}\le 26$.  This problem is particularly
relevant for the analysis of the angular correlation function since in
this statistic all galaxies at a given redshift contribute with the
same weight.  This is different, for example, to what happens when
these objects are used to estimate the star formation rate history,
where brighter objects, with smaller uncertainties in the redshift
determination, have more weight.  For these reasons we try to provide
a rough estimate of the errors in the redshift estimates at faint
magnitudes, by comparing the results of different photometric redshift
techniques and by using Monte Carlo simulations.  This in turn
provides the necessary information to define optimal redshift bin
sizes (i.e. minimizing the effects of the redshift uncertainties) for
the clustering analysis.

The plan of the paper is as follows. In Section 2, we present the
photometric database and we describe the photometric redshift
technique. In Section 3, we investigate the reliability of the
photometric redshift estimates.  In Section 4, we present the results
for the angular correlation function computed in different redshift
ranges. Section 5 is devoted to a comparison of these results with the
theoretical predictions of different cosmological models belonging to
the general class of the Cold Dark Matter scenario. Finally,
discussion and conclusions are presented in Section 6.
 
\section{The photometric redshift measurement}

\subsection{The photometric database}

As a basis for the present work, we have used the photometric
catalogue produced by FLY99 on the HDF-North using the source
extraction code SExtractor (Bertin \& Arnouts 1996).  In addition to
the four optical WFPC2 bands (Williams et al. 1996), infrared
observations in J, H and Ks bands (Dickinson et al. 1999) are
incorporated.
 
A particularly valuable feature of the FLY99 catalogue is that the
optical images are used to model spatial profiles that are fitted to
the infrared images in order to measure optimal infrared fluxes and
uncertainties.  In this way, for the large majority of the objects, an
estimate of the infrared flux is available down to the fainter
magnitudes. This is a definite advantage for the derivation of
photometric redshifts.

The analysis described below has been applied to the F300W, F450W,
F606W, F814W, J, H, Ks magnitudes of 1023 objects down to $I_{AB}
\simeq 28.5$ (here we note that the magnitude $I_{AB}$ refers directly
to the photometric catalogue given by FLY99 and not to their best fit
$I_{AB}$ reported in their photometric redshift catalogue).

\subsection{The photometric redshift technique}

Various authors have explored a number of different approaches to
estimate redshifts of galaxies from deep broad-band photometric
databases.  Empirical relations between magnitudes and/or colours and
redshifts have been calibrated using spectroscopic samples (Connolly
et al. 1995; Wang, Bahcall \& Turner 1998).  Other techniques are
based on the comparison of the observed colours of galaxies with those
expected from template SEDs, either observed (Lanzetta et al. 1996;
FLY99) or theoretical (Giallongo et al. 1998) or a combination of the
two (Sawicki, Lin \& Yee 1997; hereafter SLY97). Bayesian estimation
has also been used (Ben\'{\i}tez 1998).

\subsubsection{The synthetic spectral libraries}

The type of approach followed in the present work is based on the
comparison of observed colours with theoretical SEDs and has been
described by Giallongo et al. (1998).  Here we summarize its main
ingredients:

\begin{enumerate}
\item The SEDs are derived from the GISSEL library (Bruzual \& Charlot 1999). 
The spectral synthesis models are governed by a number of free
parameters listed in Table \ref{tgc}. The star formation rate for a
galaxy with a given age is governed by the assumed {\em e-folding}
star formation time-scale $\tau$. Several values of $\tau$ and galaxy
ages are necessary to reproduce the different observed spectral types.
We also have to assume a shape for the initial mass function (IMF).
As shown by Giallongo et al. (1998), the photometric redshift estimate
is not significantly changed by using different IMFs. Here we
restricted our analysis to a Salpeter IMF.
\item In addition to the GISSEL parameters, we have added the
internal reddening for each galaxy by applying the observed
attenuation law of local starburst galaxies derived by Calzetti,
Kinney \& Storchi-Bregmann (1994) and Calzetti (1997).  The different
values of the reddening excess are listed in Table \ref{tgc}.  We have
also included the Lyman absorption produced by the intergalactic
medium as a function of redshift in the range $0 \le z
\le 5$, following Madau (1995).
\end{enumerate}

As a result we obtained a library of $2.5 \times 10^5$ spectra, which
can be used to derive the colours as a function of redshift for all
the model galaxies with an age smaller than the Hubble time at the
given redshift (which is cosmology-dependent; the adopted cosmological
parameters are also given in Table \ref{tgc}).
         
\begin{table}
\centering
\caption[]{Parameters used for the library of templates}
\begin{tabular}{rr}
\hline 
IMF & Salpeter \\ \hline 
Exponential SFR & \\ 
Timescales $\tau$ (Gyr) & 1,2,3,5,9,$\infty$,2 bursts\\ \hline 
Ages (Gyr)& .01,.05,.1,.25,.5,.75,1.,1.5,2.,\\ \cline{2-2}
&3.,4.,5.,6.,7.,8.,9.,10.,11.,12.,14.\\ \hline 
Metallicities & ~~~~~~~~~~~~~~$Z_{\odot}$, 0.2$Z_{\odot}$, 0.02$Z_{\odot}$
\\ \hline
$E_{B-V}$ & 0,0.05,0.1,0.2,0.3,0.4 \\ \hline 
Extinction Law & Calzetti \\ \hline 
Cosmology ($H_0$, $q_0$) & 50, 0.5 \\ \hline
\end{tabular}
\label{tgc}
\end{table}

\subsubsection{Estimating redshifts}

To measure the photometric redshifts we used a standard $\chi^2$
fitting procedure comparing the observed fluxes $F_{\rm obs}$ (and
corresponding uncertainties) with the GISSEL templates $F_{\rm tem}$:
\begin{equation}
\chi ^2 = \sum_i \left[ {F_{{\rm obs},i}-s\cdot F_{{\rm tem},i} \over
\sigma_i} \right]^2 \ ,
\end{equation}
where $F_{{\rm obs},i}$ and $\sigma_i$ are the fluxes observed in a
given filter $i$ and their uncertainties, respectively; $F_{{\rm
tem},i}$ are the fluxes of the template in the same filter; the sum
runs over the seven filters. The template fluxes have been normalized
to the observed ones by choosing the factor $s$ which minimizes the
$\chi^2$ value ($\partial \chi^2 / \partial s = 0 $):
\begin{equation}
  s = \sum_i \left[ {F_{{\rm obs},i} \cdot  F_{{\rm tem},i} \over
\sigma_i^2} \right]  \Big/  \sum_i \left[ {F_{{\rm tem},i}^2 \over
\sigma_i^2} \right] \ .
\end{equation}

In the GISSEL library the models provide fluxes emitted per unit mass
(in $M_{\odot}$) and the normalization parameter $s$, which rescales
the template fluxes to the observed ones, provides a rough estimation
of the observed galaxy mass.  We have limited the range of models
accepted in the $\chi^2$ comparison to the interval $10^7$--$10^{14} \
M_{\odot}$.  We derived the $\chi^2$ probability function (CPF) as a
function of $z$ using the lowest $\chi^2$ values at any redshift.  To
have an idea of the redshift uncertainties we have derived the
interval corresponding to the standard increment $\Delta \chi^2 = 1$.
At the same time the CPF is analyzed to detect the presence, if any,
of secondary peaks with a multi-thresholding algorithm (typically we
decompose the normalized CPF into ten levels).\\ We notice that our
estimates of the photometric redshifts are changed by less than 2\% if
we adopt a different cosmology [($\Omega_{\rm 0m}=0.3$, $\Omega_{\rm
0\Lambda}=0$) or ($\Omega_{\rm 0m}=0.3$, $\Omega_{\rm 0\Lambda}=0.7$)]
and our mass estimates are nearly unchanged.
   
\section{Comparison with previous works and simulations}

\subsection{Spectroscopic vs. photometric redshifts}
\begin{figure*}
\centering
\hbox{ 
\subfigure{ \psfig{figure=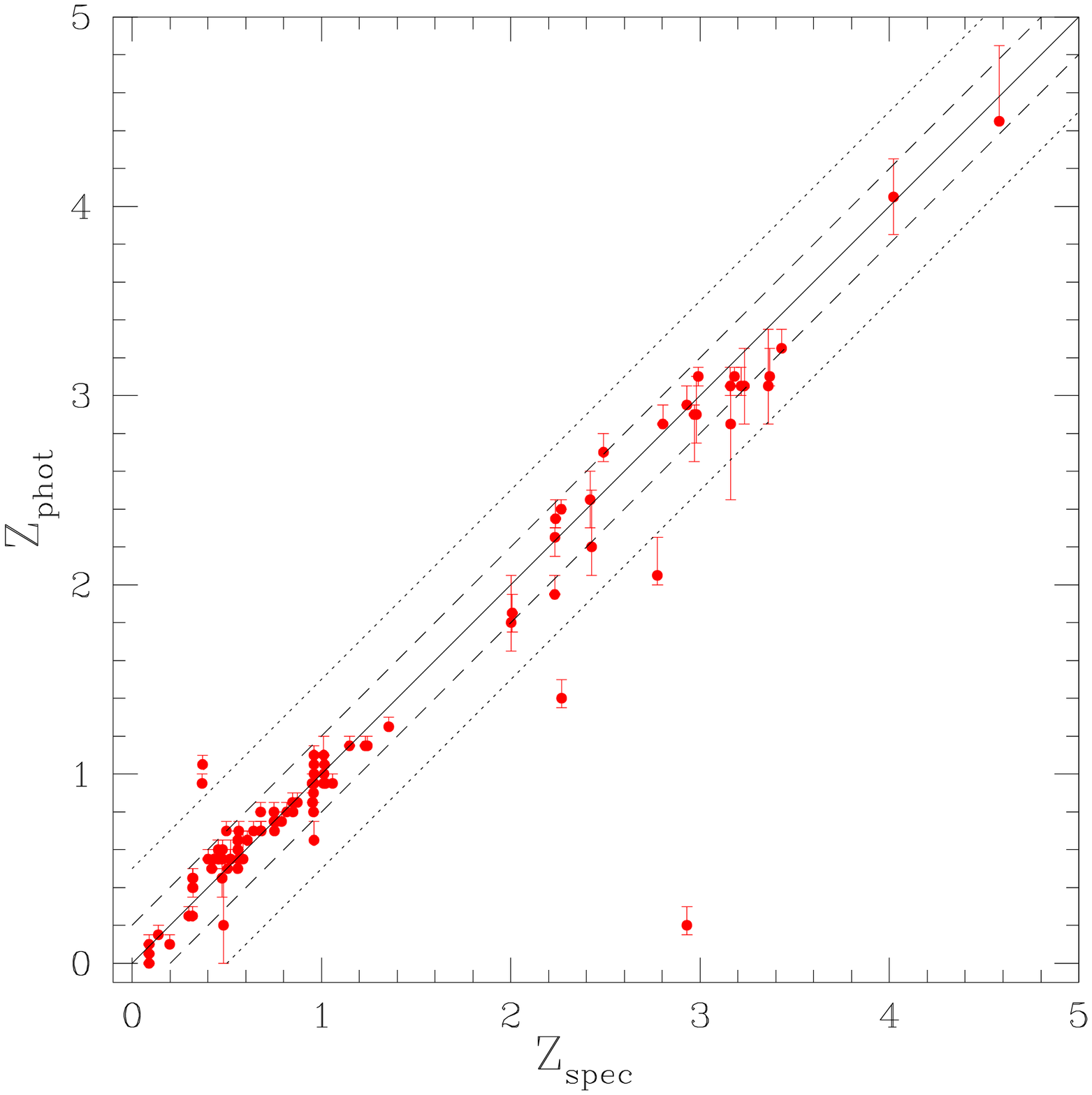,angle=0,width=7.5cm,height=7.5cm}}
\subfigure{ \psfig{figure=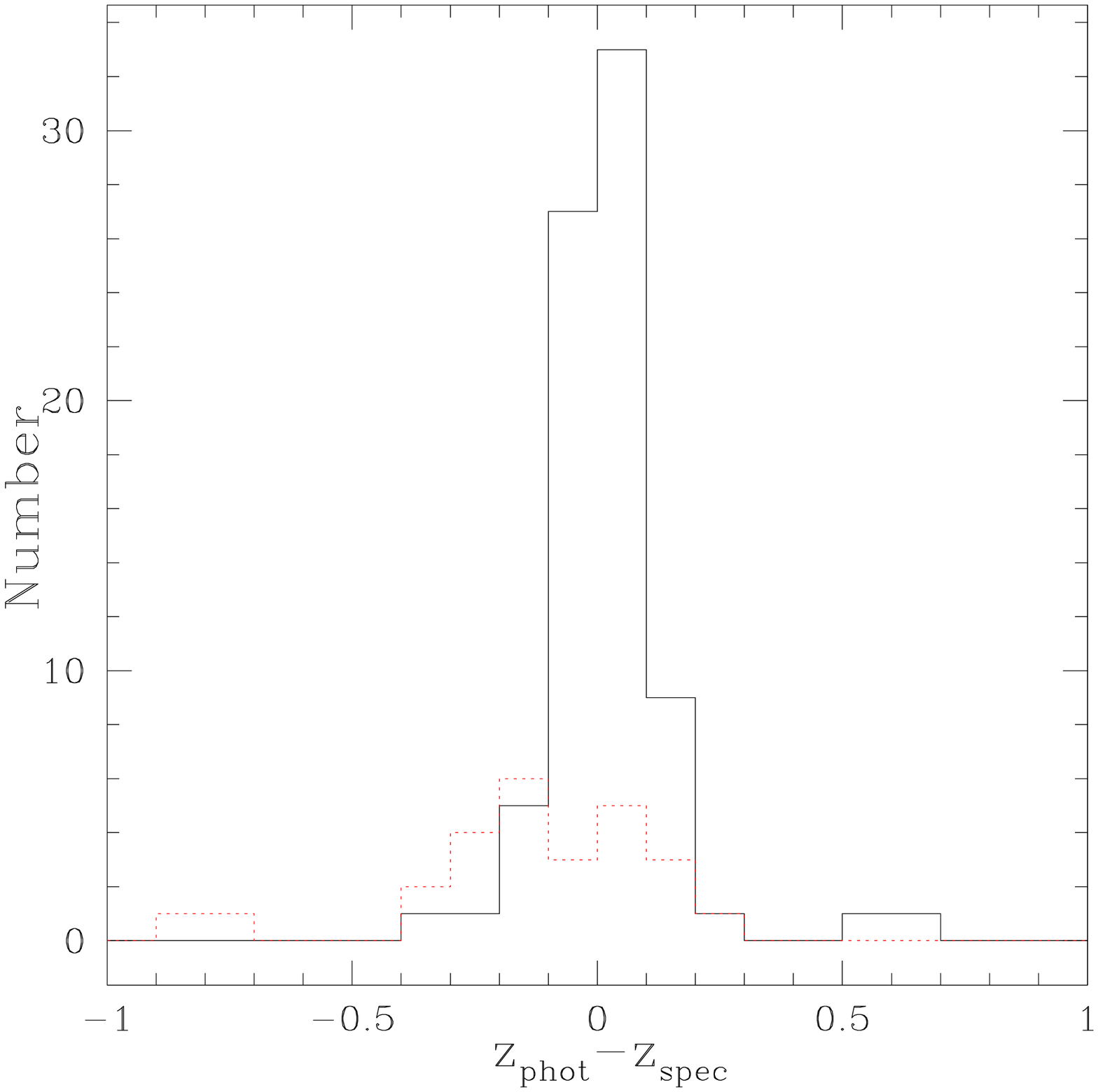,angle=0,width=7.5cm,height=7.5cm}}     }
\caption[]{The left panel shows the comparison of our photometric 
redshifts $z_{\rm phot}$ with the spectroscopic ones $z_{\rm spec}$.
Error bars represent the region where $\Delta \chi^2 \le 1$.  The
dotted and long-dashed lines represent $\Delta z = 0.5$ and $\Delta z
= 0.2$, respectively.  The right panel shows the histograms with
$z_{\rm phot} - z_{\rm spec}$ for galaxies with $z_{\rm spec}\le 1.5$
(solid line) and $z_{\rm spec} > 1.5$ (dotted line).  }
\label{fzszp1} \end{figure*}

In Figure~\ref{fzszp1} we show the comparison of our estimates of the
photometric redshifts $z_{\rm phot}$ with the 106 spectroscopic
redshifts $z_{\rm spec}$ up to $z\simeq 5$ listed in the FLY99
catalogue (see references therein).  Our values are generally
consistent with the observed spectroscopic redshifts within the
estimated uncertainties over the full redshift range.  The
r.m.s. dispersion $\sigma_z$ for different redshift intervals is
reported in Table~\ref{tzcomp}.  At redshifts lower than 1.5, two
galaxies have photometric redshifts which appear clearly discrepant:
galaxy \# 191 (the number refers to the FLY99 number) with $z_{\rm
phot} \simeq 1.05$ vs. $z_{\rm spec}\simeq 0.37$ and galaxy \# 619
with $z_{\rm phot} \simeq 0.95$ vs. $z_{\rm spec} \simeq 0.37$.  Also
FLY99 and SLY97 found for these two objects $z_{\rm phot} \ge 0.88$.
As discussed in the next section, the techniques used in SLY97, in
FLY99 and in the present work are significantly different; therefore,
if the spectroscopic redshifts are correct, both objects are expected
to have a really peculiar SED.  For example, various SEDs used in
these works do not include spectra with strong emission lines
(starbursts, AGN, ...). Yet, based on the observed spectra, the two
spectroscopic redshifts are very uncertain (see {\tt
http://astro.berkeley.edu/davisgrp/HDF/}).  Disregarding these two
objects, the photometric accuracy at $z<1.5$ decreases from $\sigma_z
\simeq 0.13$ to $\sigma_z \simeq 0.09$.  These values are consistent
with the photometric redshift estimates obtained in previous works and
compiled by Hogg et al. (1998).

At redshifts $z\ge 1.5$ the dispersion is $\sigma_z = 0.24$, if the
galaxy \# 687, which shows catastrophic disagreement (it is found at
low redshift also by FLY99, while there is no clear association in the
SLY97 catalogue) is discarded.  Direct inspection of the original
frames shows that in this case the photometry can be incorrect due to
the complex morphology of this object, which was assumed to be a
single unit.

\begin{table}
\centering
\caption[]{ Comparison of our photometric redshifts with 106
spectroscopic redshifts up to $z=5$ for different redshift intervals
(Column 1).  Here we consider only objects either with $|\Delta z|
\equiv | z_{\rm spec}-z_{\rm phot} | \le 1$ or with $| \Delta z | \le
0.5$ (Column 2).  The corresponding number of objects in each redshift
interval is given in Column 3 and the associated dispersion $\sigma_z$ in 
Column 4. }
\begin{tabular}{cccc}
\hline
  $z$ range  & $| \Delta z | $  & $N_{\rm phot}$/$N_{\rm spec}$  &  
$\sigma_z$  \\
\hline
0.0  -  5.0     &   $\le$ 1.0     & 105/106  & 0.20   \\
0.0  -  1.5     &   $\le$ 1.0     & 79/79    & 0.13  \\
1.5  -  5.0     &   $\le$ 1.0     & 28/29    & 0.24 \\
0.0  -  5.0     &   $\le$ 0.5     & 101/106  & 0.12 \\
0.0  -  1.5     &   $\le$ 0.5     & 77/79    & 0.09  \\
1.5  -  5.0     &   $\le$ 0.5     & 26/29    & 0.15 \\
\hline
\end{tabular}
\label{tzcomp}
\end{table}

%
\subsection{Comparison with other photometric redshifts }
The relatively good agreement of the photometric redshifts with the
spectroscopic ones shows the reliability of our method at bright
magnitudes.  Obviously, the same accuracy cannot be expected also at
fainter magnitudes, below the spectroscopic limit $I_{AB}\ge 26$.  The
uncertainty in the identification of the characteristic features (4000
\AA \ and Lyman break) in the observed SEDs necessarily increases when
the errors in the photometry become larger.  In order to obtain a
rough idea of the uncertainty also in the domain inaccessible to
spectroscopy we have compared the results of our code with those
obtained with other photometric methods.

FLY99 and SLY97 have used the four spectra provided by Coleman, Wu \&
Weedman (1980) which reproduce different star formation histories or
different galaxy types (E/S0, Sbc, Scd and Irr).  The wavelength
coverage of these template spectra is however too small (1400 - 10000
\AA) to allow a direct comparison with the full range of photometric
data (3000 - 25000 \AA).  To bypass this problem, both authors have
extrapolated the infrared SEDs by using the theoretical SEDs of the
GISSEL library, corresponding to the four spectral types. In the UV
SLY97 have used again an extrapolation based on GISSEL while FLY99
have used the observations of Kinney et al. (1993).  SLY97 have
enlarged the SED library with two spectra of young galaxies with a
constant star formation (from the GISSEL library) and interpolated
between the six spectra to reduce the aliasing effect due to the SED
sparse sampling.

The comparison of the two approaches with spectroscopic redshifts has
been carried out by the authors: the uncertainties are typically
$\sigma_z \simeq 0.10-0.15$ at $z \le 1.5$ and reach $\sigma_z \simeq
0.20-0.25$ at higher redshifts.

\begin{figure*}
\centering
\vbox{
\hbox{ 
\subfigure{ \psfig{figure=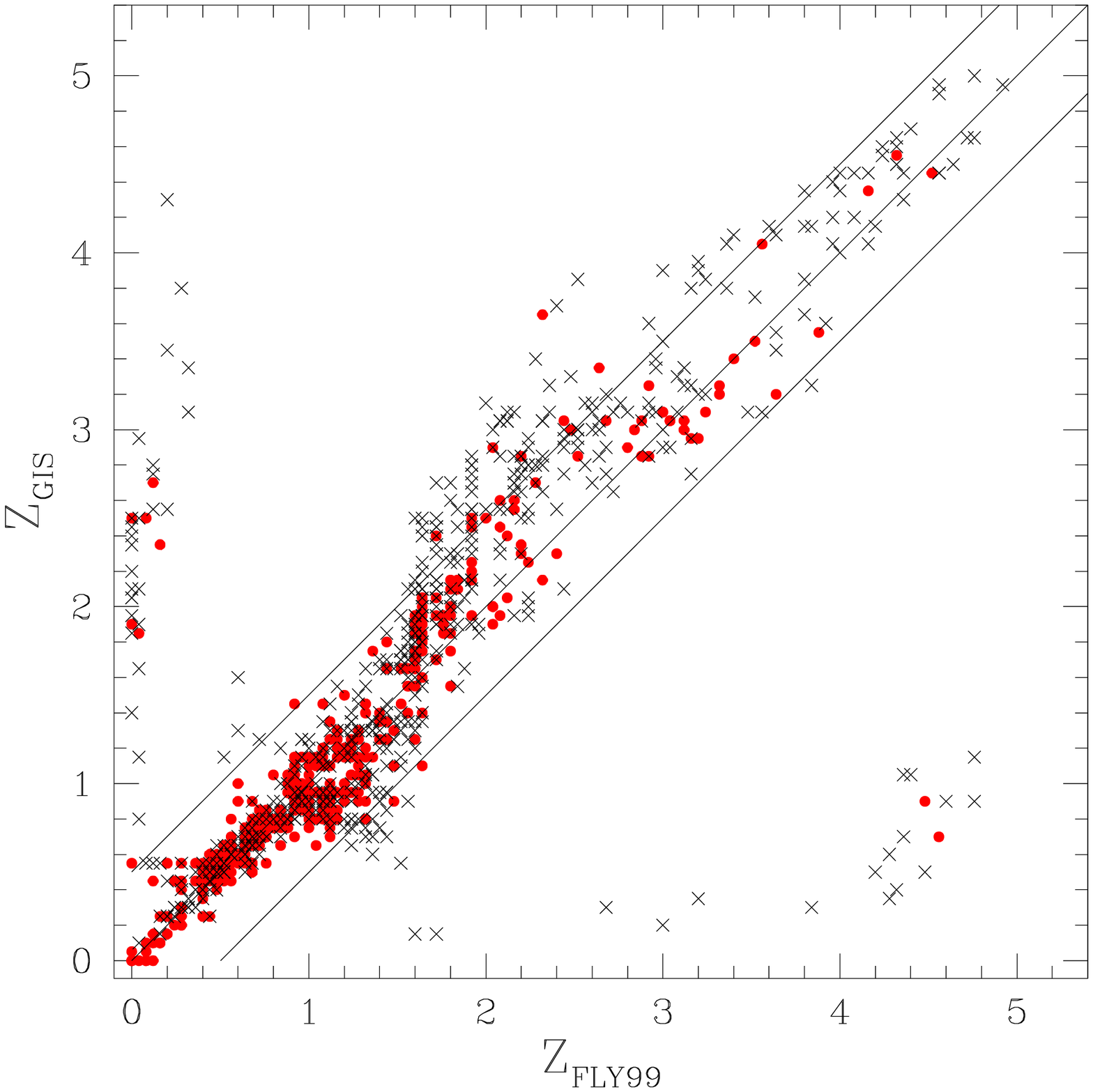,angle=0,width=7.5cm,height=7.5cm}}
\subfigure{ \psfig{figure=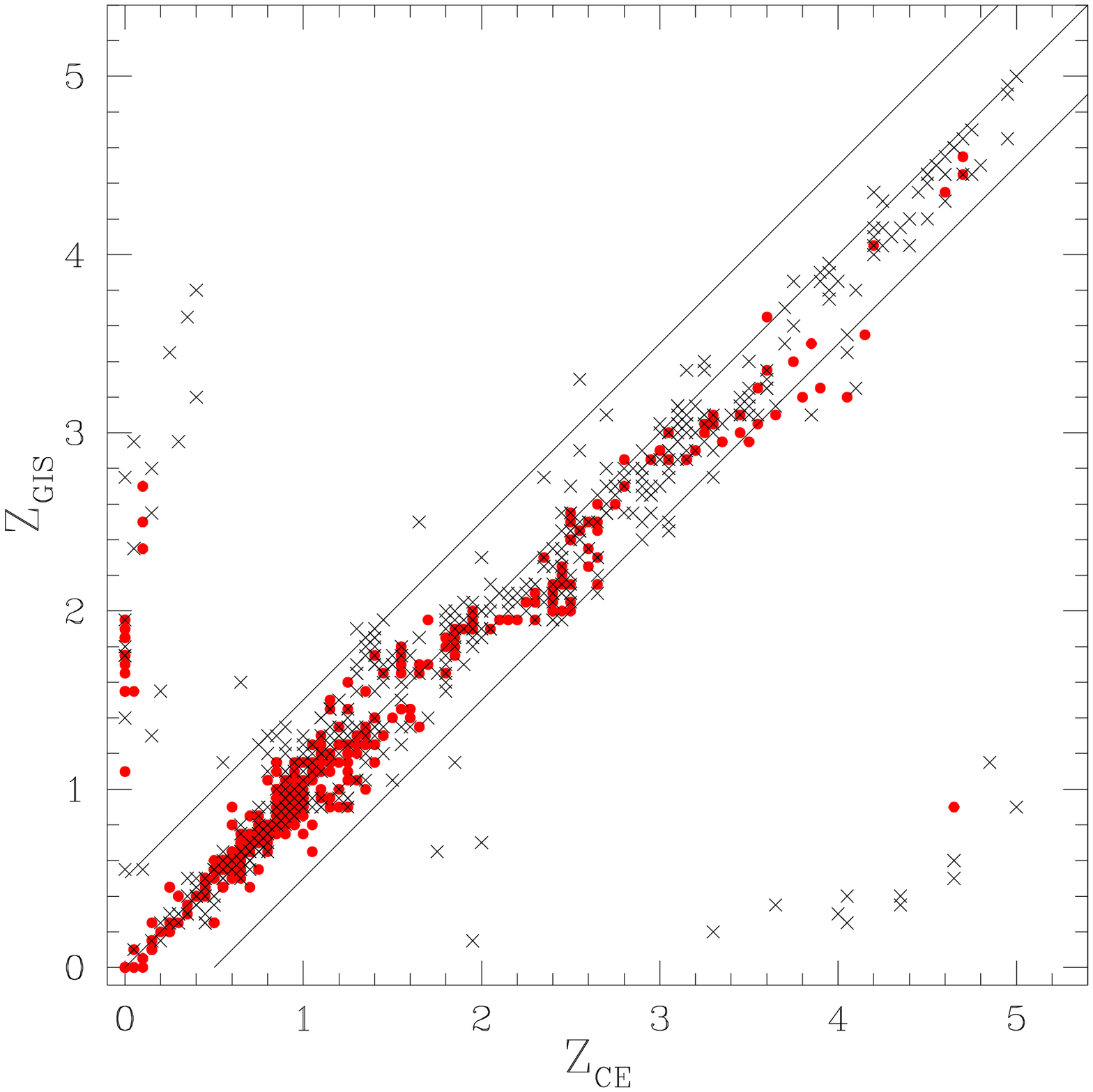,angle=0,width=7.5cm,height=7.5cm}} 
     }
\hbox{ 
\subfigure{ \psfig{figure=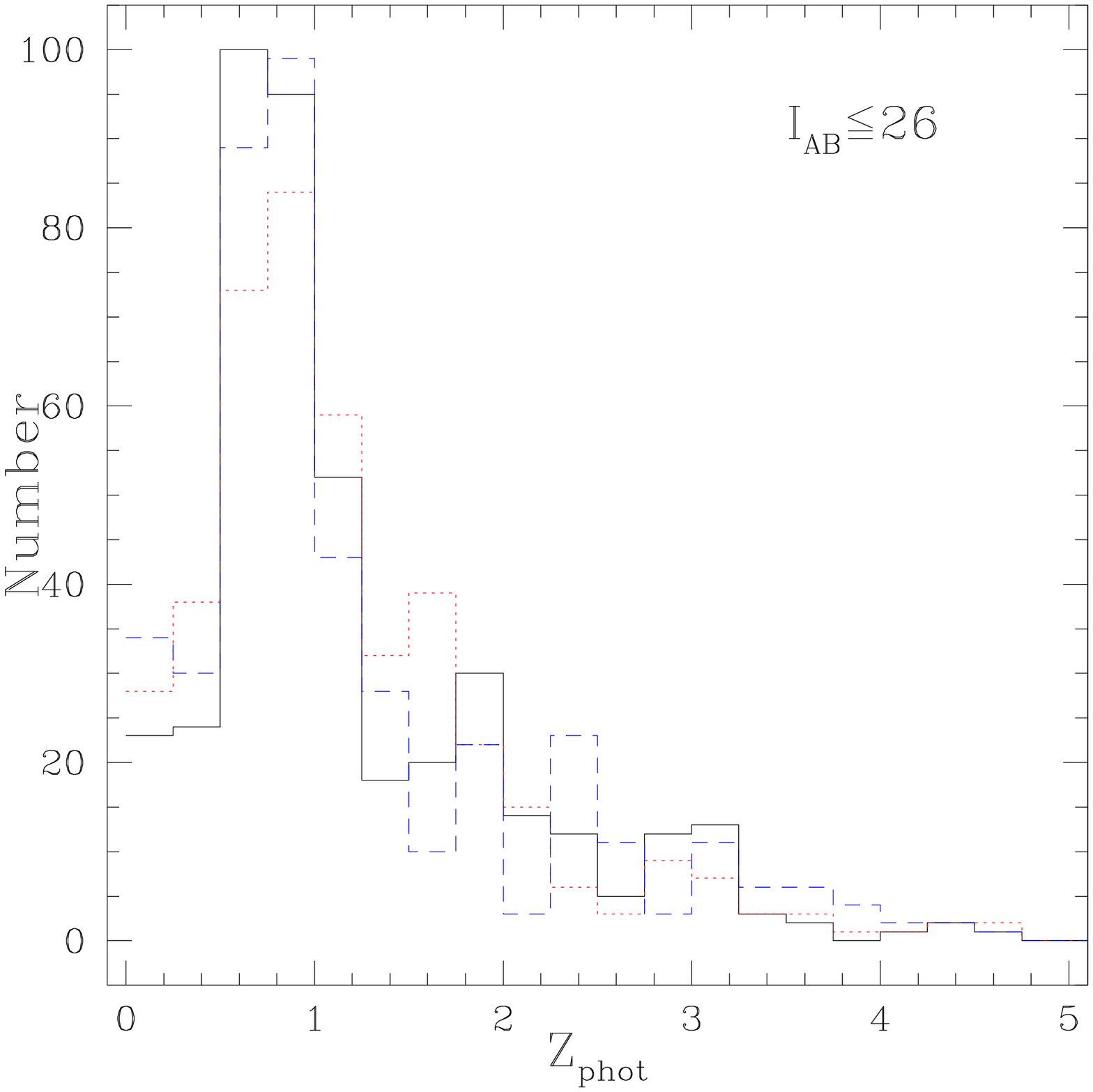,angle=0,width=7.5cm,height=7.5cm}}
\subfigure{ \psfig{figure=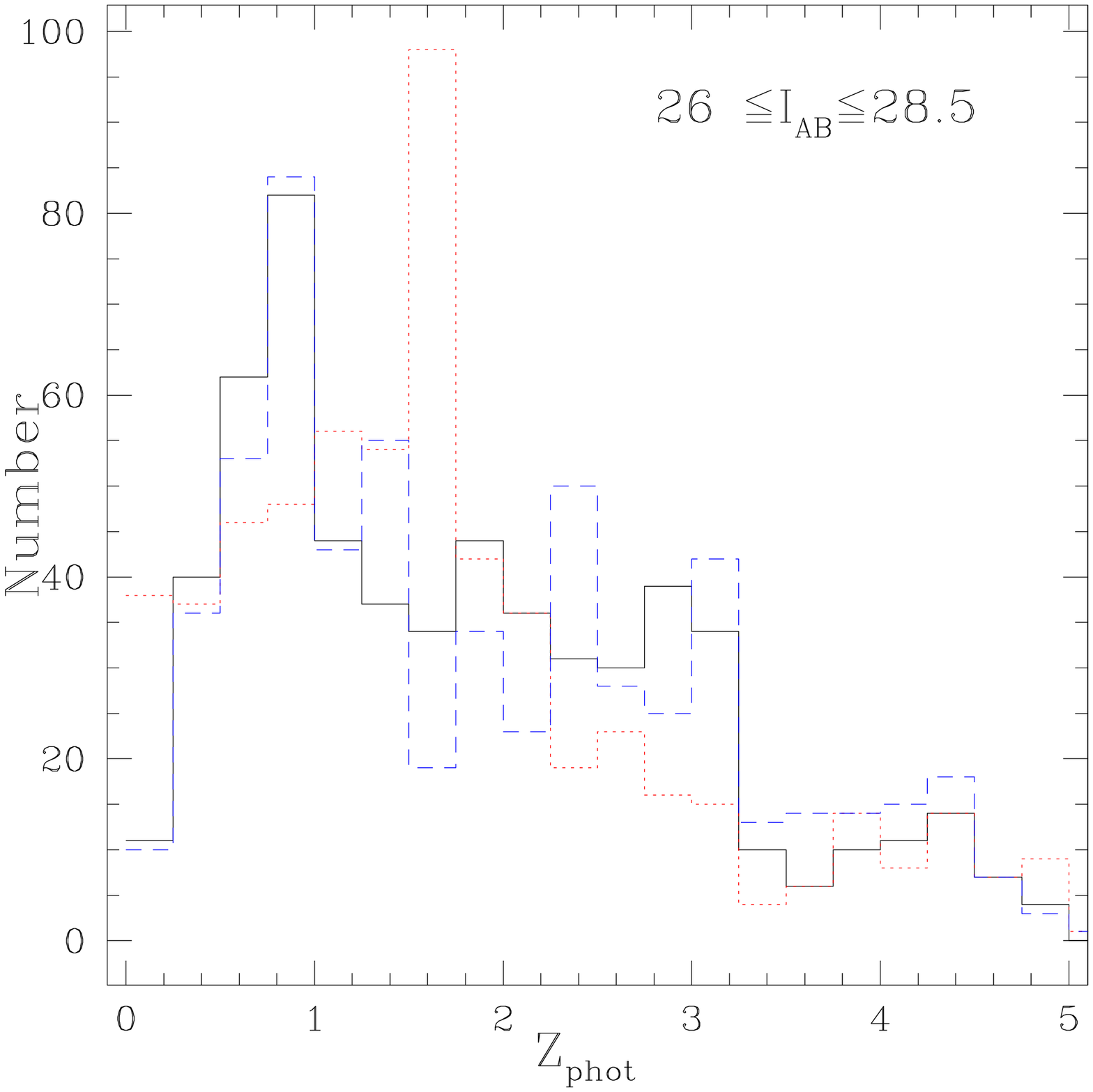,angle=0,width=7.5cm,height=7.5cm}} 
     }
     }
\caption[]{Comparison of our photometric redshifts $z_{\rm GIS}$ (computed
using the GISSEL library) with the photometric redshifts obtained by
FLY99 ($z_{\rm FLY99}$; upper left panel) and with our Coleman
Extended (CE) libraries similar to the SLY97 method ($z_{\rm CE}$;
upper right panel).  Filled circles represent objects with $I_{AB}\le
26 $ and crosses refer to objects with $ 26 \le I_{AB}\le 28.5$.
Solid lines correspond to $\Delta z = 0.5$. In the two lower panels,
we show the comparison of the three redshift histograms for $I_{AB}\le
26$ (left panel) and $ 26 < I_{AB}\le 28.5$ (right panel).  Our GISSEL
model, the CE model and the FLY99 model are shown by solid, dashed and
dotted lines, respectively.  }
\label{fzpcomp} \end{figure*}

In the SLY97 analysis, only the four optical bands have been used to
estimate the photometric redshifts.  To carry out a fair comparison,
we have set up a code based on a library similar to that used by SLY97
and we recomputed the photometric redshifts with the FLY99 catalogue
(hereafter called Coleman Extended model: CE).  The comparison between
the three methods is shown in Figure~\ref{fzpcomp} (upper panels).
The three redshift distributions are shown in the lower panels of the
the same figure.  From these plots, we observe that:
\begin{enumerate}
\item For $I_{AB}\le 26$ the three methods are compatible 
within $\Delta z \simeq 0.5$.  A small number ($\sim$ 2\%) of
catastrophic discrepancies ($\Delta z \ge 1$) is observed.  Excluding
these objects, we find r.m.s. dispersions $\sigma_z\simeq 0.12$ and
$\sigma_z\simeq 0.23$ between the GISSEL and CE models at $z\le 1.5$
and $1.5<z\le 5$, respectively. In the high-redshift range a
systematic shift is observed with $\langle z_{\rm GIS} - z_{\rm CE}
\rangle \simeq -0.15$.  Between the GISSEL and FLY99 models, the
dispersions are $\sigma_z\simeq 0.16$ and $\sigma_z\simeq 0.26$ at
$z\le 1.5$ and $1.5<z\le 5$, respectively, with a systematic shift in
the high-redshift range $\langle z_{\rm GIS} - z_{\rm FLY99} \rangle
\simeq +0.18$.  These results are compatible with the uncertainties
based on the spectroscopic sample.  Finally the three resulting
redshift distributions are in good agreement.
\item  For $I_{AB}\le 28.5$ the number of objects with $\Delta z \ge 1$  
increases and represents the 6\% of the full sample in both cases.
Excluding these objects, we find dispersions $\sigma_z\simeq 0.18$ and
$\sigma_z\simeq 0.26$ between the GISSEL and CE models at $z\le 1.5$
and $1.5<z\le 5$, respectively.  For the high-redshift range a
systematic shift is still observed with $\langle z_{\rm GIS} - z_{\rm
CE} \rangle \simeq -0.11$. Comparing the GISSEL and FLY99 models, the
dispersions are $\sigma_z\simeq 0.22$ and $\sigma_z\simeq 0.32$ at
$z\le 1.5$ and $1.5<z\le 5$, respectively, with a larger systematic
shift in the high-redshift range $\langle z_{\rm GIS} - z_{\rm FLY99}
\rangle \simeq +0.31$.
\item  The large shift for $z\ge 1.5$ observed with FLY99 is due  
to a feature appearing in their redshift distribution with a large
number of sources between $1.2 \le z \le 2$, not observed in the two
other models (Figure~\ref{fzpcomp}, lower right panel).  The interval
$1.2 \le z \le 2$ is critical for the photometric determination of the
redshifts, due to the lack of strong features.  In fact the
Lyman-alpha break is not yet observed in the $F300W$ band and the
break at $4000$ \AA\ is located between the $F814W$ and $J$ bands.
Therefore the estimates rest basically on the continuum shape of the
templates. As shown by FLY99 in their Figure 6, their photometric
redshifts suffer from a systematic underestimate with respect to the
spectroscopic ones around $z\simeq 2$. This may be due to an
inadequacy of the UV extrapolation used by FLY99 in reproducing the UV
shape of the high-$z$ objects.  This effect disappears at higher
redshifts because of the U-dropout effect.  As a check, we have added
to the four templates of FLY99 a spectrum of an irregular galaxy with
constant star formation rate (with higher UV flux).  In this case the
excess of galaxies with $1.2\le z \le 2$ disappears and the objects
are re-distributed in better agreement with the two other methods.
\item  Our GISSEL model produces a smaller  number of objects  at $z\ge 3.5$ 
with respect to the two other approaches. The discrepant objects
(found at lower redshift by the GISSEL code) are generally fitted by
using a significant fraction of reddening excess ($\langle {\rm
E(B-V)} \rangle \simeq 0.3$). Note that in general objects found at
$z\ge 3.5$ by the GISSEL code are also at high redshift with the other
techniques.
\end{enumerate} 
%
%
%
\subsection{Comparison with the NICMOS F110W and F160W observations}
\begin{figure}
\centering  
\psfig{figure=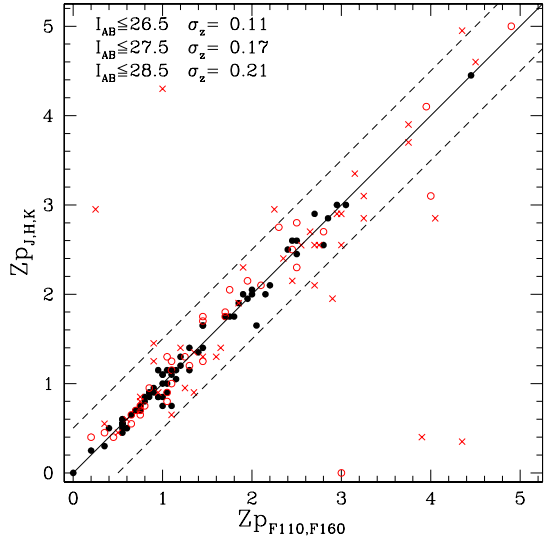,height=7.5cm,width=7.5cm,angle=0}
\caption{Comparison of the photometric redshifts obtained by using our
GISSEL model and those obtained by replacing the J, H, Ks filters with
the $F110W$ and $F160W$ filters for 164 objects. Filled circles
represent galaxies with $I_{AB}\le 26.5$, open circles represent
galaxies with $26.5\le I_{AB}\le 27.5$ and crosses refer to galaxies
with $27.5\le I_{AB}\le 28.5$.  The dashed lines represent $\Delta z =
0.5$. The redshift dispersions $\sigma_z$ for different magnitude
limits are given inside the figure.  }
\label{fzpnic}
\end{figure}

Recently, deep NICMOS images have been obtained in the area
corresponding to chip 4 of the WFPC2 camera in the HDF-North (Thompson
et al. 1999).  The observations have been carried out in the two
filters $F110W$ and $F160W$ and reach $F160W_{AB} \simeq 28.8$ (at
3$\sigma$). We have associated each NICMOS detection (from the
published catalogue) with the FLY99 catalogue.  We consider in our
analysis the 164 objects detected in both NICMOS filters.  These data
provide a crucial check thanks to their depth and high spatial
resolution and also to the spectral coverage of the $F110W$ band. This
filter fills the gap between the $F814W$ filter and the standard $J$
filter and makes it possible to detect the 4000 \AA \ break at $z\ge
1.2$.  We have recomputed the photometric redshifts with our GISSEL
models using the four optical bands and replacing the J, H, Ks filters
with the $F110W$ and $F160W$ filters.  The results are shown in
Figure~\ref{fzpnic}.  This subsample shows a good agreement between
NICMOS and J, H, Ks photometry and corroborates the reliability of the
infrared measurements performed by FLY99.  The redshift agreement in
the range $0\le z \le 5$ is better than $|\Delta z|=0.5$ up to
magnitudes $I_{AB}\le 28.5$ and only 5/164 objects present
discrepancies with $|\Delta z| \ge 1$.

%
\subsection{Comparison with Monte Carlo simulations}
\begin{figure*}
\centering  
\psfig{figure=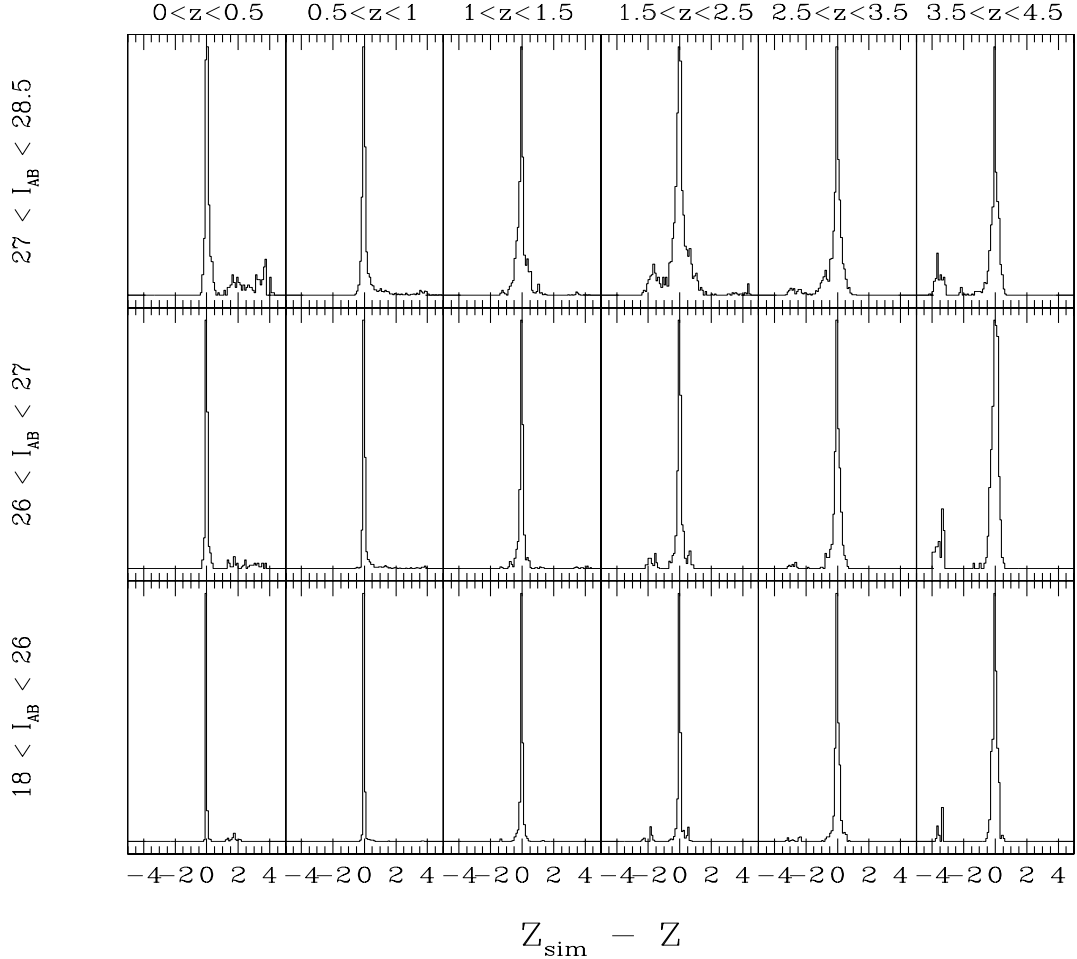,height=13.cm,width=15.cm,angle=0}
\caption{Effect of the photometric errors in the redshift estimates.
We have built a catalogue of approximately 21,000 simulated galaxies
by adding to the fluxes of each original object a gaussian random
noise with r.m.s. equal to the flux uncertainties in each band.  The
histograms of the differences ($\Delta z=z_{\rm sim}-z$) between the
simulated redshift $z_{\rm sim}$ and the original one $z$ is shown for
different magnitude and redshift ranges.  }
\label{fzpsim}
\end{figure*}

As final check we performed Monte Carlo simulations to study the
effect of photometric errors on our redshift estimates.  To do so we
have added to the original fluxes of the 1067 galaxies of the FLY99
catalogue a gaussian random noise with r.m.s. equal to the flux
uncertainties in each band.  This operation has been repeated 20 times
to produce a catalogue of approximately 21,000 simulated galaxies for
which we have re-estimated the photometric redshifts with our code. In
Figure~\ref{fzpsim}, we show the distribution of the differences
$\Delta z$ between the simulated redshifts $z_{\rm sim}$ and the
original ones $z$ ($\Delta z = z_{\rm sim} - z$) for different
magnitude and redshift ranges.  Several comments can be made from this
figure.
\begin{enumerate}
\item  The median value of the redshift difference is very 
close to zero ($\le 0.05$) for any magnitude and redshift range. The
dispersion around the peak, $\sigma_z$, is larger for larger
magnitudes and redshifts. In Table~\ref{tzcont} we report $\sigma_z$
for galaxies with $I_{AB}\le 28.5$ for different redshift
ranges. These dispersions are compatible with the observed ones based
on the comparison made above between different codes.
\item 
Table~\ref{tzcont} also reports the number of simulated galaxies put
in a redshift bin different from their original one because of the
photometric errors (Column 3).  These results show that the number of
lost original galaxies varies between 15\% to 25\% at any redshift for
$I_{AB}\le 28.5$.  In the redshift range $0\le z\le 0.5$, the
discrepant objects are distributed in a high redshift tail between $1
\le z \le 4$.  For the three bins with $z\ge 1.5$, the discordant
objects are preferentially located in a secondary peak at low $z$
($0\le z_{\rm sim} \le 1$).
\item  The galaxies 
lost from an original bin are a contaminating factor for the others.
We can estimate for each bin this contamination which is also reported
in Table~\ref{tzcont} (Column 4). In the same table the contaminating
fraction due only to the adjacent bins is reported (Column 5).  We can
see that the contamination plays a different role at different
redshifts.  For $0\le z\le 0.5$, the contamination is quite large
($\simeq 30\%$) and it is not due to the adjacent bin (representing
only one third of the total). In this case the main source of
contamination are high-redshift galaxies put at low redshifts. For the
other bins the contamination is close to 20\% and is essentially due
to the adjacent bins.
\end{enumerate}
     
%
\begin{table}
\caption[]{Contamination effects for different redshift intervals
computed from Monte Carlo simulations. Column 1 indicates the redshift
range.  Column 2 reports the dispersions $\sigma_z$ around the higher
peak in the distribution of the redshift differences $\Delta z$ for
simulated galaxies up to $I_{AB}=28.5$.  Column 3 shows the fractions
of objects which are outside the original redshift bin (Lost).  Column
4 reports the contamination by objects belonging to another original
redshift bin (Cont.).  Finally Column 5 reports the contamination by
objects belonging to the original adjacent redshift bins
(Adj. Cont.).}

\begin{tabular}{ccccc}
\hline
  $z$ range&   &Simulations&$I_{AB}<28.5$ & \\
\hline
    &$\sigma_z$&Lost &Cont. &Adj. Cont.\\
    &         & (\%) & (\%) &  (\%)   \\
\hline
 0.0 - 0.5 &  0.20 & 19.3 & 30.2 & 9.4 \\
 0.5 - 1.0 &  0.20 & 12.2 & 11.5 & 9.3\\
 1.0 - 1.5 &  0.25 & 25.0 & 15.5 & 12.5\\
 1.5 - 2.5 &  0.35 & 22.7 & 27.0 & 22.7\\
 2.5 - 3.5 &  0.32 & 22.1 & 19.2 & 16.9\\
 3.5 - 4.5 &  0.26 & 26.3 & 21.8 & 15.1\\
\hline
\end{tabular}
\label{tzcont}
\end{table}

\section{The angular correlation function}    
\subsection{Definition of the redshift 
bin sizes and subsamples}

We have limited our analysis to the region of the HDF with the highest
signal-to-noise, excluding the area of the PC, the outer part of the
three WFPC and the inner regions corresponding to the junction between
each chip. In this area we included in our sample all galaxies
brighter than $I_{AB}\simeq 28.5$. This procedure leads to a slight
reduction of the overall number of galaxies: our final sample contains
959 out of the 1023 original ones.
 
To correctly compute the angular correlation function (ACF) the
following details have to be taken into account:
\begin{enumerate}
\item the relatively small field of view of the HDF (the angular 
distance corresponds to $\simeq 1 h^{-1}$ Mpc at $z\ge 1$, with
$q_0=0.5$);
\item  the accuracy  of the photometric redshifts;
\item  the number of objects in each redshift bin, in order to
reduce the shot noise and achieve sufficient sensitivity to the
clustering signal.
\end{enumerate} 
As a consequence, relatively large redshift bins are required:
according to Figure~\ref{fzpcomp} and Table~\ref{tzcont}, a minimum
redshift bin size of $\Delta z = 0.5$ (corresponding to $\Delta z \sim
2 \times \sigma_z$) is required for $z\le 1.5$.  At higher redshifts,
due to the uncertainties in the redshifts and the relatively low
surface densities, a more appropriate bin size is $\Delta z = 1$.
Moreover, these large bin sizes can reduce the effects of redshift
distortion and, most important, attenuate the sample variance effect
caused by the small area covered by the HDF North (approximately 4
arcmin$^2$). A refined approach to treat the sample variance has been
recently proposed by Colombi, Szapudi \& Szalay (1998).

Finally, we note that the contamination discussed in the previous
section can introduce a dilution of the clustering signal. In the
worst case, assuming that the contaminating population is
uncorrelated, it introduces a dilution of about $(1-f)^2$ (where $f$
corresponds to the contaminating fraction reported in
Table~\ref{tzcont}). This correction factor has been used to define
upper-limits to the clustering estimates which are shown in the
following figures.

\subsection{The computation of the Angular Correlation Function}
 
The angular correlation function $\omega(\theta)$ is related to the
excess of galaxy pairs in two solid angles separated by the angle
$\theta$ with respect to a random Poisson distribution.  The angular
separation used for the computation of $\omega(\theta)$ covers the
range from 5 arcsec up to 80 arcsec. We use logarithmic bins with
steps of $\Delta \log \theta = 0.3$.  The lower limit makes it
possible to avoid a spurious signal at small scales due to the
multi-deblending of resolved bright spirals and irregulars, the upper
cut-off is almost half the size of the HDF and corresponds to the
maximum separation where the ACF provides a reliable signal.

\begin{figure*}
\centerline{\psfig{figure=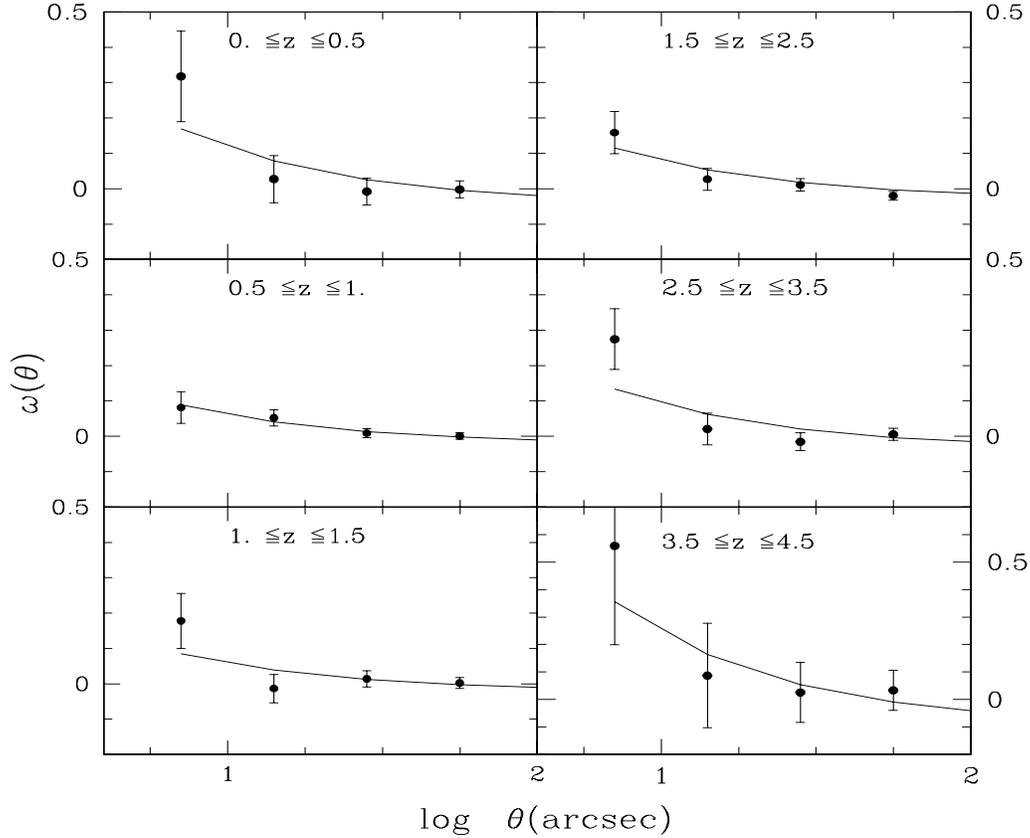,angle=0,width=14.cm,height=12.cm}}
       \caption[]{ The angular correlation functions $\omega(\theta)$,
       computed with the estimator of Landy \& Szalay (1993), for
       galaxies with $I_{AB} \le 28.5$ measured for different redshift
       ranges (as specified in each panel). The uncertainties are
       Poisson errors.  The solid lines show the best fits obtained by
       assuming $\omega(\theta) = A_{\omega}\ \theta^{-\delta}$, with
       a fixed slope $\delta = 0.8$.  }
\label{fwdz} \end{figure*}

To derive the ACF in each redshift interval, we used the estimator
defined by Landy \& Szalay (1993):
\begin{equation}
 \omega_{\rm est}(\theta) \ = \frac{DD(\theta) - 2\ DR(\theta) +
 RR(\theta) } {RR(\theta)}\ ,
\end{equation}    
where DD is the number of distinct galaxy-galaxy pairs, DR is the
number of galaxy-random pairs and RR refers to random-random pairs
with separation between $\theta$ and $\theta + \Delta \theta$.  The
random catalogue contains 20,000 sources covering the same area of our
sample.  In Figure~\ref{fwdz} we show the measured ACF for each
redshift bin. The uncertainties are Poisson errors as shown by Landy
\& Szalay (1993) for this estimator.

Adopting a power-law form for the ACF as $\omega(\theta)=A_{\omega}
\theta^{-\delta}$, we derive the amplitude $A_{\omega}$ assuming
$\delta= \gamma - 1 = 0.8$. Here $\gamma$ is the slope of the spatial
correlation function, which is also assumed to follow a power-law
relation.  Formally, we can use both $A_{\omega}$ and $\delta$ as free
parameters to be obtained from the least-square fitting, but due to
the limited sample, we prefer to fix $\delta$ and leave as free
parameter only $A_{\omega}$.  The value of the slope we assume is
larger than the estimates obtained by Le F\`evre et al. (1996) in the
analysis of the CFRS catalogue (which covers the interval $0 \leq z
\leq 1$), and is smaller than the estimates obtained for LBGs by G98
at $z\simeq3$.  Nevertheless, the adopted value is still consistent
with the respective uncertainties. The value of the slope could also
depend on the magnitude, as discussed by Postman et al. (1998).

To estimate the amplitude of the ACF, due to the small size of the
field, we introduce the integral constraint IC in our fitting
procedure as $\omega_{\rm est} \simeq \omega_{\rm true} - IC =
A_{\omega} \times ( \theta^{-0.8} - B)$.  The quantity $IC =
A_{\omega} \times B$ has been computed by a Monte Carlo method using
the same geometry of the HDF and masking the excluded regions.  In
this computation, we adopt the same value for the slope ($\delta=0.8$)
and we derive $B = 0.044$ (for $\theta$ measured in arcsec).  The best
fits for the ACF in each redshift bin are shown as solid lines in
Figure~\ref{fwdz}.  The amplitudes $A_{\omega}$ obtained by the best
fits are listed in Table~\ref{tawz} with the adopted magnitude limits
and the number of galaxies used.  We give also the measured amplitude
for the galaxies with $I_{AB}\le 28.5$ and with $0 \le z \le 6$.  All
these values are not corrected for the contamination factor.
  
    \begin{figure*}
\centering
\psfig{figure=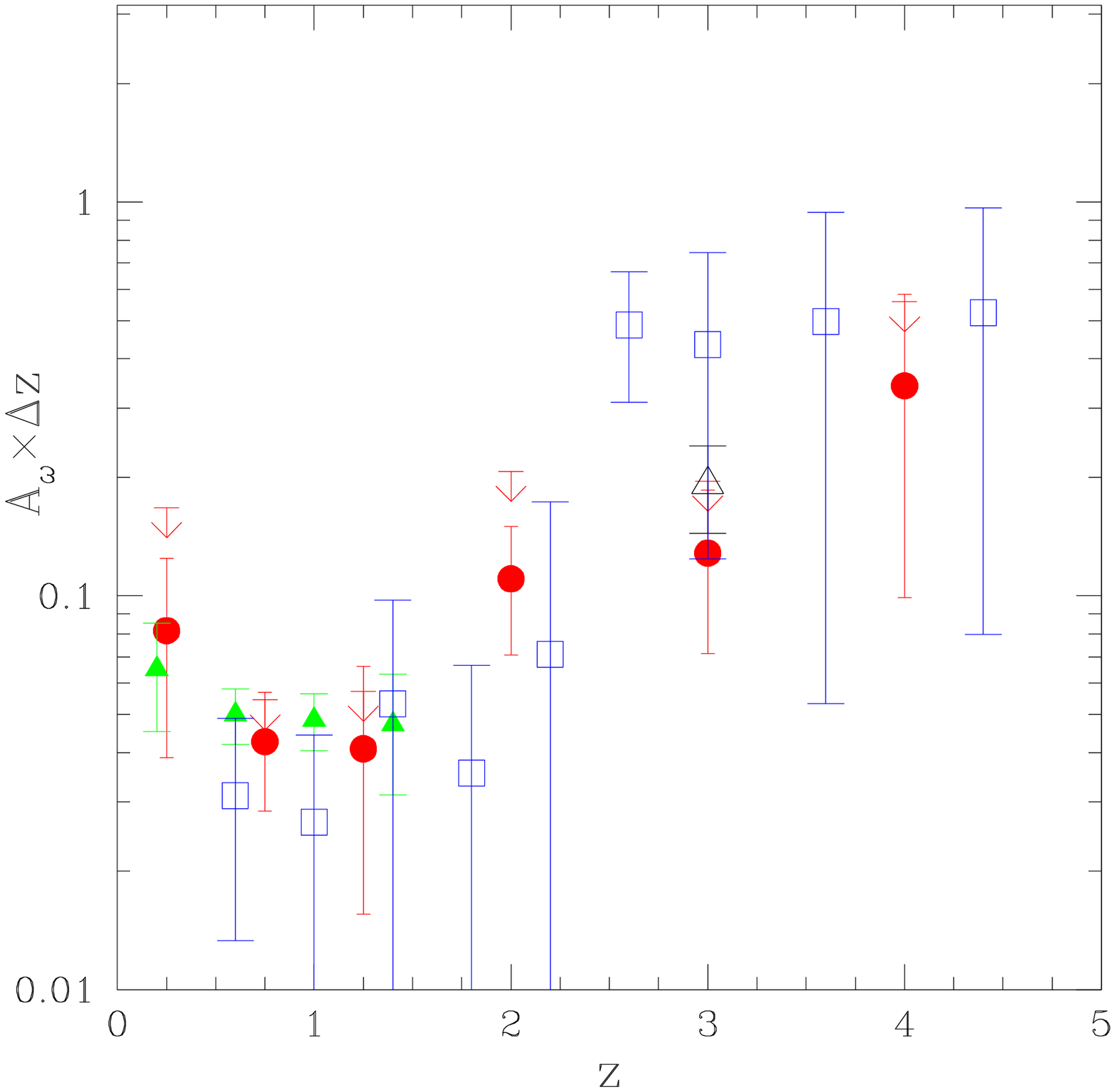,angle=0,width=8.cm,height=8.cm}
\caption[]{ 
The amplitude of $\omega(\theta)$ at 10 arcsec ($A_{\omega}$) as a
function of redshift. The values are rescaled to the same $\Delta z$
by applying $A_{\omega} \times \Delta z$ for a direct comparison (see
text). Filled circles represent our results for $I_{AB}\le 28.5$.  The
arrows show our measurements corrected for the contamination factor
and should be considered as upper-limits.  The filled triangles show
the values obtained by Connolly et al. (1998). The open triangle is
the value for the LBG sample (Giavalisco et al. 1998) and open squares
refer to the values obtained by Magliocchetti \& Maddox (1999).  }
\label{fawz}
     \end{figure*}

In Figure~\ref{fawz} we compare our values of $A_{\omega}$ (at 10
arcsec) to other published data (Connolly et al. 1998; G98;
Magliocchetti \& Maddox 1999). 
The values of $A_{\omega}$ take into account the adopted redshift 
bin sizes.  At a given redshift, a larger $\Delta z$ implies smaller
$A_{\omega}$ due to the increasing number of foreground and background
galaxies with respect to the unchanged number of physically correlated
pairs ($A_{\omega} \propto \Delta z^{-1}$; see e.g. Connolly et
al. 1998).  Then, if we assume that $A_{\omega}$ does not strongly
evolve inside the redshift bin, we can correct the original amplitudes
by using $A_{\omega} \times \Delta z$, which allows a more direct
comparison.  From this figure we note that our results are in good
agreement with those of Connolly et al. (1998) and slightly smaller
than those for LBGs obtained by G98. The agreement with Magliocchetti
\& Maddox (1999) is worse but still consistent with both
error bars. In this figure we show the possible effect of the
contamination factor discussed in the previous section. This
correction increases all the values which should be regarded as
upper-limits due to the basic assumption that the contaminating
population is uncorrelated.  Moreover we notice that our estimate in
the redshift bin $0\le z \le 0.5$ can be affected by the lack of
nearby bright galaxies in the HDF. For this reason, this point will be
not considered in the following comparison between observational
results and model predictions

\section{Comparison with theoretical models}

\subsection{The formalism}

We can now predict the behaviour of the angular correlation function
$\omega(\theta)$ for our galaxy sample in various cosmological
structure formation models. The angular two-point function for a
sample extended in the redshift direction over an interval ${\cal Z}$
can be written in terms of the spatial correlation function using the
relativistic Limber equation (Peebles 1980).  We adopt here the Limber
formula as given in Matarrese et al. (1997), namely
\be
\omega_{\rm obs}(\theta) = N^{-2} \int_{\cal Z} 
dz \biggl({dr \over dz}\biggr)^{-1} ~{\cal N}^2(z) 
\int_{-\infty}^{\infty}  du ~\xi_{\rm gal} [r(u,\theta,z),z] \;
\ee 
where $r(u,\theta,z)=\sqrt{u^2 + r^2(z)\theta^2}$, in the small-angle
approximation (e.g. Peebles 1980).

The relation between the comoving radial coordinate $r$ and the
redshift $z$ is given with whole generality by
\be
r(z) =  {c \over H_0 \sqrt{|\Omega_{0\cal R}|}}     
{\cal S} \left(\sqrt{|\Omega_{0\cal R}|}
\int_0^z \left[\left( 1+z' \right)^2 \left(1+\Omega_{\rm 0m} z'\right) -
z'\left(2+z'\right) \Omega_{0\Lambda}\right]^{-1/2} dz'\right)   \;, 
\label{eq:x_z}
\ee
where $\Omega_{0\cal R} \equiv 1 - \Omega_{\rm 0m} -
\Omega_{0\Lambda}$, with $\Omega_{\rm 0m}$ and $\Omega_{0\Lambda}$ the
density parameters for the non-relativistic matter and cosmological
constant components, respectively. In this formula, for an open
universe model, $\Omega_{0\cal R}>0$, $S(x)\equiv \sinh (x)$, for a
closed universe, $\Omega_{0\cal R}<0$, $S(x)\equiv \sin (x)$, while in
the EdS case, $\Omega_{0\cal R} = 0$, $S(x) \equiv x$.

In the Limber equation above, ${\cal N}(z)$ is the redshift
distribution of the catalogue (whose integral over the entire redshift
interval is $N$), which is given by ${\cal N}(z) = \int_{\cal M} d \ln
M {\cal N}(z,M)$, with ${\cal N}(z,M) = 4\pi g_c(z) \phi(z,M) \bar
n_c(z,M)$ and $\bar n_c(z,M)$ is the expected number of galaxies per
comoving volume at redshift $z$; $\phi(z,M)$ is the isotropic
catalogue selection function. The quantity ${\cal N}(z,M)$ represents
the number of objects actually present in the catalogue, with redshift
in the range $z,~z+dz$ and intrinsic properties (like mass,
luminosity, ...) in the range $M,~M+dM$ (${\cal M}$ representing the
overall interval of variation of $M$).  In the latter integral we also
defined the comoving Jacobian
\be
g_c(z) \equiv  r^2(z) 
\biggl[1 + {H_0^2 \over c^2} ~\Omega_{0\cal R} ~r^2(z) \biggr]^{-1/2} 
{dr \over dz} \;. 
\ee

In what follows we will assume a simple model for our galaxy
distribution, where galaxies are associated in a one-to-one
correspondence to their hosting dark matter haloes. The advantage of
this model is that haloes can be simply characterized by their mass
$M$ and formation redshift $z_f$. Since haloes merge continuously into
larger mass ones one can safely assume that their formation redshift
coincides with the observation one, namely $z_f=z$.  This simple model
of galaxy clustering was named `transient' model in Matarrese et
al. (1997) and Moscardini et al. (1998); Coles et al. (1998) adopted
it to describe the clustering of LBGs. The application of this model
is more appropriate at high redshifts where merging dominates while at
low redshifts it can only be a rough approximation.  Recently Baugh et
al. (1999) showed that this simple model under-predicts the clustering
properties at low redshift because it does not take into account the
possibility that a single halo can host more than one galaxy.  Indeed,
as discussed in Moscardini et al. (1998), a `galaxy conserving' bias
model is likely to provide a better description of the galaxy
clustering evolution at low redshift.

In practice, in our modelling we select a minimum mass $M_{\rm min}$
for the haloes hosting our galaxies, i.e. we take
$\phi(z,M)=\theta(M-M_{\rm min})$, with $\theta$ the Heaviside step
function, and we compute the corresponding value of the effective bias
$b_{\rm eff}$ (see equation below) at each redshift.  In what follows
we will consider two possibilities: {\em i}) $M_{\rm min}$ fixed to a
sensible value (we will show results obtained by using $10^{10}$,
$10^{11}$ and $10^{12}$ $h^{-1} M_\odot$), {\em ii}) $M_{\rm min} =
M_{\rm min}(z)$ chosen to reproduce a relevant set of observational
data.  For the latter case we will adopt two different strategies: in
the first case we assume $M_{\rm min}(z)$ so that the theoretical
${\cal N}(z)$ fits the observed one in each redshift bin (e.g. Mo \&
Fukugita 1996; Moscardini et al. 1998; A98; Mo, Mao \& White 1999); in
the second case we adopt at any redshift the median of the mass
distribution estimated by our GISSEL model. Actually this model gives
a rough estimate of the baryonic mass. To convert it to the mass of
the hosting dark matter halo we multiply by a factor 10.  This value
corresponds to a baryonic fraction close to that predicted by the
standard theory of primordial nucleosynthesis. Variations in the range
from 5 to 20 produce only small changes in the following results.

As a first, though accurate, approximation the galaxy spatial
two-point function can be taken as being linearly proportional to that
of the mass, namely $\xi_{\rm gal}(r,z) \simeq b_{\rm eff}^2(z)
\xi_{\rm m}(r,z)$, where  
\be 
b_{\rm eff}(z) \equiv {\cal N}(z)^{-1} \int_{\cal M} d\ln M' 
~{\cal N}(z,M') ~b(M',z) 
\label{eq:b_eff}
\ee
is the effective bias of our galaxy sample and $\xi_{\rm m}$ the
matter covariance function.

The bias parameter $b(M,z)$ for haloes of mass $M$ at redshift $z$ in
a given cosmological model can be modeled as (Mo \& White 1996)
\be
b(M,z) = 1 + {1 \over \delta_c }
\biggl( {\delta_c^2 \over \sigma_M^2 D_+^2(z) } - 1\biggr) \;, 
\label{eq:b_mono}
\ee
where $\sigma^2_M$ is the linear mass-variance averaged over the scale
$M$, extrapolated to the present time ($z=0$), $\delta_c$ the critical
linear overdensity for spherical collapse ($\delta_c={\rm
const}=1.686$ in the EdS case, while it depends slightly on $z$ for
more general cosmologies) and $D_+(z)$ is the linear growth factor of
density fluctuations (e.g. $D_+(z) = (1+z)^{-1}$ in the EdS case).  In
comparing our theoretical predictions on clustering with the data, we
will always adopt for the galaxy redshift distribution ${\cal N}(z)$
the observed one. Nevertheless, consistency requires that the
predicted halo redshift distribution for a given minimum halo mass
always exceeds (because of the effects of the selection function) the
observed galaxy one.  For the calculation of the effective bias, where
we need ${\cal N}(z,M)$, one might adopt the Press \& Schechter (1974)
recipe to compute the comoving halo number density (per unit
logarithmic interval of mass); it reads
\be
\bar n_c(z,M) =  \sqrt{2 \over \pi} {{\bar \varrho_0} \delta_c 
\over 
M  D_+(z) \sigma_M } \bigg| {d \ln \sigma_M 
\over d \ln M} \bigg| 
   \exp \biggl[ -{\delta_c^2  \over 2 D_+^2(z) \sigma^2_M}
\biggr] \;
\label{eq:ps}
\ee
(with $\bar \varrho_0$ the mean mass density of the Universe at $z=0$).

However, a number of authors have recently shown that the
Press-Schechter formula does not provide an accurate description of
the halo abundance both in the large and small-mass tails (see
e.g. the discussion in Sheth \& Tormen 1999). Also, the simple Mo \&
White (1996) bias formula of Equation (\ref{eq:b_eff}) has been shown
not to correctly reproduce the correlation of low mass haloes in
numerical simulations. Several alternative fits have been recently
proposed (Jing 1998; Porciani, Catelan \& Lacey 1999; Sheth \& Tormen
1999; Jing 1999). An accurate description of the abundance and
clustering properties of the dark matter haloes corresponding to our
galaxy population will be obtained here by adopting the relations
introduced by Sheth \& Tormen (1999), which have been obtained by
fitting to the distribution of the halo population of the GIF
simulations (Kauffmann et al. 1999): this technique allows to
simultaneously improve the performance of both the mass function and
the bias factor. The relevant formulas, replacing
Eqs.(\ref{eq:b_mono}) and (\ref{eq:ps}) above, read
\be
b(M,z) =   1 + {1 \over \delta_c }
\biggl( {a \delta_c^2 \over \sigma_M^2 D_+^2(z)} - 1\biggr) 
          + {2 p  \over \delta_c }
\biggl( {1 \over { 1+[\sqrt{a} \delta_c / (\sigma_M D_+(z))]^{2p}}}
\biggr) 
\;  
\label{eq:b_mono2}
\ee 
and
\be 
\bar n_c(z,M) =  \sqrt{2 a A^2 \over \pi} {{\bar \varrho_0} \delta_c 
\over M  D_+(z) \sigma_M } 
\biggl[ 1 + \biggl( {{D_+(z) \sigma_M} \over {\sqrt{a}  \delta_c}}
\biggr)^{2p} \biggr] 
  \bigg| {d \ln \sigma_M 
\over d \ln M} \bigg| \exp \biggl[ -{a \delta_c^2  \over 
2 D_+^2(z) \sigma^2_M}
\biggr] \; ,
\label{eq:ps2}
\ee 
respectively. In these formulas $a=0.707$, $p=0.3$ and
$A\approx0.3222$, while one would recover the standard (Mo \& White
and Press \& Schechter) relations for $a=1$, $p=0$ and $A=1/2$.

The computation of the clustering properties of any class of objects
is completed by the specification of the matter covariance function
$\xi_{\rm m}(r,z)$ and its redshift evolution. To this purpose we
follow Matarrese et al. (1997) and Moscardini et al. (1998) who used
an accurate method, based on the Hamilton et al. (1991) original
ansatz to evolve $\xi_{\rm m}(r,z)$ into the fully non-linear
regime. Specifically, we use here the fitting formulas proposed by
Peacock \& Dodds (1996).

As recently pointed out by various authors (e.g. Villumsen 1996;
Moessner, Jain \& Villumsen 1998), when the redshift distribution of
faint galaxies is estimated by applying an apparent magnitude limit
criterion, magnification bias due to weak gravitational lensing would
modify the relation between the intrinsic galaxy spatial correlation
function and the observed angular one. Modelling this effect within
the present scheme would be highly desirable, but is certainly beyond
the scope of our work. Nevertheless, we note that this magnification
bias would generally lead to an increase of the apparent clustering of
high-$z$ objects above that produced by the intrinsic galaxy
correlations, by an amount which depends on the amplitude of the
fluctuations of the underlying matter distribution.

\subsection{Structure formation models}

We will consider here a set of cosmological models belonging to the
general class of Cold Dark Matter (CDM) scenarios.  The linear
power-spectrum for these models can be represented by $P_{\rm
lin}(k,0) \propto k^n T^2(k)$, where we use the fit for the CDM
transfer function $T(k)$ given by Bardeen et al. (1986), with ``shape
parameter" $\Gamma$ defined as in Sugiyama (1995).  To fix the
amplitude of the power spectrum (generally parameterized in terms of
$\sigma_8$, the r.m.s. fluctuation amplitude inside a sphere of $8
h^{-1}$ Mpc) we either attempt to fit the local cluster abundance,
following the Eke, Cole \& Frenk (1996) analysis of the temperature
distribution of X-ray clusters (Henry \& Arnaud 1991), or the level of
fluctuations observed by COBE (Bunn \& White 1997). In particular, we
consider the following models: A version of the standard CDM (SCDM)
model with $\sigma_8=0.52$, which reproduces the local cluster
abundance, but is inconsistent with COBE data.  The so-called
$\tau$CDM model (White, Gelmini \& Silk 1995), with shape parameter
$\Gamma=0.21$. A COBE normalized tilted model, hereafter called TCDM
(Lucchin \& Matarrese 1985), with $n=0.8$, $\sigma_8=0.52$ and high
(10 per cent) baryonic content (e.g. White et al.  1996; Gheller,
Pantano \& Moscardini 1998); the normalization of the scalar
perturbations, which takes into account the production of
gravitational waves predicted by inflationary theories (e.g. Lucchin,
Matarrese \& Mollerach 1992; Lidsey \& Coles 1992), allows to
simultaneously fit the CMB fluctuations observed by COBE and the local
cluster abundance.  The three above models are all flat and without
cosmological constant.  We also consider here: A cluster normalized
open CDM model (OCDM), with matter density parameter $\Omega_{\rm
0m}=0.3$, and $\sigma_8=0.87$, which is also consistent with COBE
data. Finally, a cluster normalized low-density CDM model
($\Lambda$CDM), with $\Omega_{\rm 0m}=0.3$, but with a flat geometry
provided by the cosmological constant, with $\sigma_8=0.93$, which is
also consistent with COBE data. A summary of the parameters of the
cosmological models used here is given in Table \ref{t:models}.

\begin{table}
\centering
\caption[]{The parameters of the cosmological models. Column 2: the present
matter density parameter $\Omega_{\rm 0m}$; Column 3: the present
cosmological constant contribution to the density $\Omega_{0\Lambda}$;
Column 4: the primordial spectral index $n$; Column 5: the Hubble
parameter $h$; Column 6: the shape parameter $\Gamma$; Column 7: the
spectrum normalization $\sigma_8$.}
\tabcolsep 4pt
\begin{tabular}{lcccccc} \\ \\ \hline \hline
Model & $\Omega_{\rm 0m}$ & $\Omega_{0\Lambda}$ & $n$ & $h$ &
$\Gamma$ & $\sigma_8$  \\ \hline
SCDM         & 1.0 & 0.0 & 1.0 & 0.50 & 0.45 & 0.52  \\
$\tau$CDM    & 1.0 & 0.0 & 1.0 & 0.50 & 0.21 & 0.52  \\
TCDM         & 1.0 & 0.0 & 0.8 & 0.50 & 0.41 & 0.52  \\
OCDM         & 0.3 & 0.0 & 1.0 & 0.65 & 0.21 & 0.87  \\
$\Lambda$CDM & 0.3 & 0.7 & 1.0 & 0.65 & 0.21 & 0.93  \\
\hline
\end{tabular}
\label{t:models}
\end{table}

\subsection{Results}

\begin{figure*}
\centering
\psfig{figure=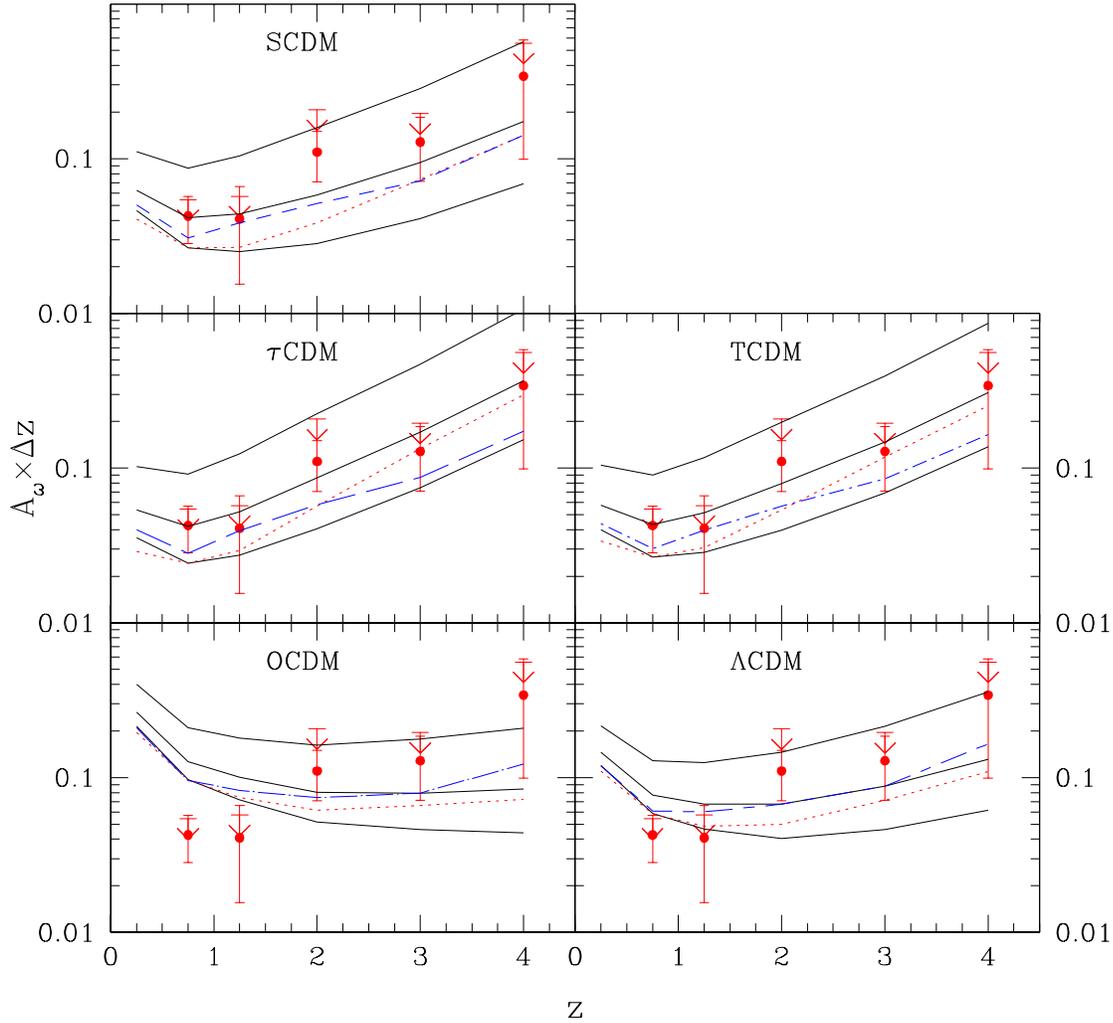,angle=0,width=15.cm,height=15.cm}
        \caption[]{Comparison of the observed $A_{\omega}$ (filled
        circles for $I_{AB}\le 28.5$, arrows for the upper-limits
        estimates) with the prediction of the various theoretical
        models described in the text. The solid lines show the
        measurements expected when a minimum mass $M_{\rm min} =
        10^{10}$, $10^{11}$ and $10^{12} h^{-1} M_{\odot}$ is assumed;
        the lower curves refer to smaller masses. The dotted lines
        show the prediction obtained by using the median masses at any
        redshift estimated by our GISSEL model shown in
        Figure~\ref{fawmass} (baryonic masses are translated into the
        masses of the hosting dark matter haloes by multiplying by a
        factor of 10; see text). The dashed curves correspond to
        models where the masses necessary to reproduce the observed
        density of objects in each redshift bin are used. The dashed
        curve types used for the different cosmological models are the
        same as those used in Figure~\ref{fawmass}.}
\label{fawmod} \end{figure*}

In Figure~\ref{fawmod} we compare the observed amplitude of the ACF
with the predictions of the various cosmological models. For
consistency with the analysis performed on the observational data
shown in the previous section, here the theoretical results have been
obtained by fitting the data in the same range of angular separation
and using the same stepping $\Delta \log \theta=0.3$. A fixed slope of
$\delta=0.8$ is also used in the following analysis. Notice that this
value is only a rough estimate of the best fit slopes: generally the
resulting values are smaller ($\delta \simeq 0.6$) in all redshift
intervals and for all the models. The discrepancy is higher for TCDM
and $\tau$CDM ($\delta \simeq 0.3-0.4$) and can lead to some ambiguity
in the interpretation of the results (see the discussion on the
effective bias below).

\begin{figure*}
\centering
\psfig{figure=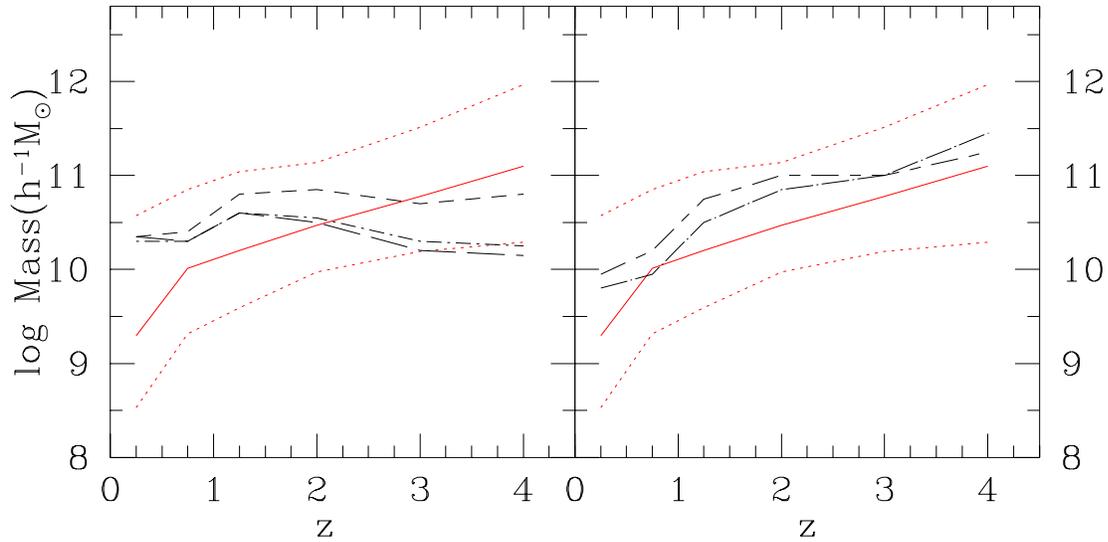,angle=0,width=15.cm,height=8.cm}
        \caption[]{In both panels, the solid line shows, as a function
        of the redshift, the median of the distribution of the galaxy
        masses estimated by our GISSEL model; the lower and upper
        quartiles are shown by dotted lines.  The baryonic masses are
        translated into the masses of the hosting dark matter haloes
        multiplying by a factor 10. We show also the mass necessary to
        reproduce the observed density of objects at any $z$ for SCDM
        (short-dashed line), $\tau$CDM (long-dashed line) and TCDM
        (short-dashed -- dotted line) models in the left panel and for
        OCDM (long-dashed -- dotted line) and $\Lambda$CDM
        (long-dashed -- short--dashed line) models in the right
        panel.}
\label{fawmass}
 \end{figure*}

In each panel the solid lines show the results obtained when we use
different (but constant in redshift) values of $M_{\rm min}$
($10^{10}$, $10^{11}$ and $10^{12}$ $h^{-1} M_\odot$ from bottom to
top). These results can be regarded as a reference on what is the
minimum mass of the galaxies necessary to reproduce the observed
clustering strength.  However, the assumption that the catalogue
samples at any redshift the same class of objects, i.e. with the same
typical minimum mass, cannot be realistic. In fact, we expect that at
high redshifts the sample tends to select more luminous, and on
average more massive, objects than at low redshifts. This is supported
by the distribution of the galaxy masses inferred by the GISSEL model,
shown in Figure~\ref{fawmass}. The solid line, which represents the
median mass, is an increasing function of redshift: from $z\simeq 0$
to $z\simeq 4$ its value changes by at least a factor of 30. In
Figure~\ref{fawmass} we also show the masses necessary to reproduce at
any redshift the observed galaxy density. In general, they are
compatible with the GISSEL distribution but the redshift dependence is
different for the various cosmological models considered here. For EdS
universe models (left panel) the different curves are quite similar
and almost constant with typical values of $10^{10.5} h^{-1}
M_\odot$. On the contrary for OCDM and $\Lambda$CDM models (shown in
the right panel) $M_{\rm min}(z)$ is an increasing function of
redshift: at $z\simeq 0$ $M_{\rm min}\simeq 10^{10} h^{-1} M_\odot$,
while at $z\simeq 4$ $M_{\rm min}\simeq 10^{11.5} h^{-1} M_\odot$. The
amplitudes of the ACF obtained by adopting these $M_{\rm min}(z)$
values are also shown in Figure~\ref{fawmod}.

In general, all the models are able to reproduce the qualitative
behaviour of the observed clustering amplitudes, i.e. a decrease from
$z=0$ to $z\simeq 1-1.5$ and an increase at higher redshifts. The EdS
models are in rough agreement with the observational results when a
minimum mass of $10^{11} h^{-1} M_\odot$ is used at any redshift. As
discussed above this mass is slightly larger than the one required to
fit the observed ${\cal N}(z)$. The situation for OCDM and
$\Lambda$CDM models is different.  The amount of clustering measured
would require that the involved objects have, at redshifts $z\le
1-1.5$, minimum masses smaller than $10^{10} h^{-1} M_\odot$, at
redshifts $1.5 \mincir z \mincir 3$, minimum masses of the order of
$10^{11.5} h^{-1} M_\odot$, while, at $z\simeq 4$, $M_{\rm min}\geq
10^{12} h^{-1} M_\odot$ is needed to reproduce the clustering
strength.  These small values at low redshifts are probably due to the
kind of biasing model adopted in an epoch when merging starts to be
less important. This is particularly true for open models and flat
models with a large cosmological constant, where the growth of
perturbations is frozen by the rapid expansion of the universe. On the
contrary, the need to explain the high amplitude of clustering at
$z\simeq 4$ with very massive objects can be in conflict with the
observed abundance of galaxies at this redshift, which requires
smaller minimum masses.

If the spatial correlation function can be written in the simple form
$\xi_{\rm gal}(r,z)= [r/r_0(z)]^{-\gamma}$, it is possible to obtain
the comoving correlation length $r_0(z)$ and the r.m.s. galaxy density
fluctuation $\sigma^{\rm gal}_8(z)$, with the assumption that the
clustering does not strongly evolve inside each redshift bin used for
the amplitude measurements (see Magliocchetti \& Maddox 1999 for the
relevant formulas in the framework of different cosmological models).

\begin{figure}
\centering
\hbox{ 
\psfig{figure=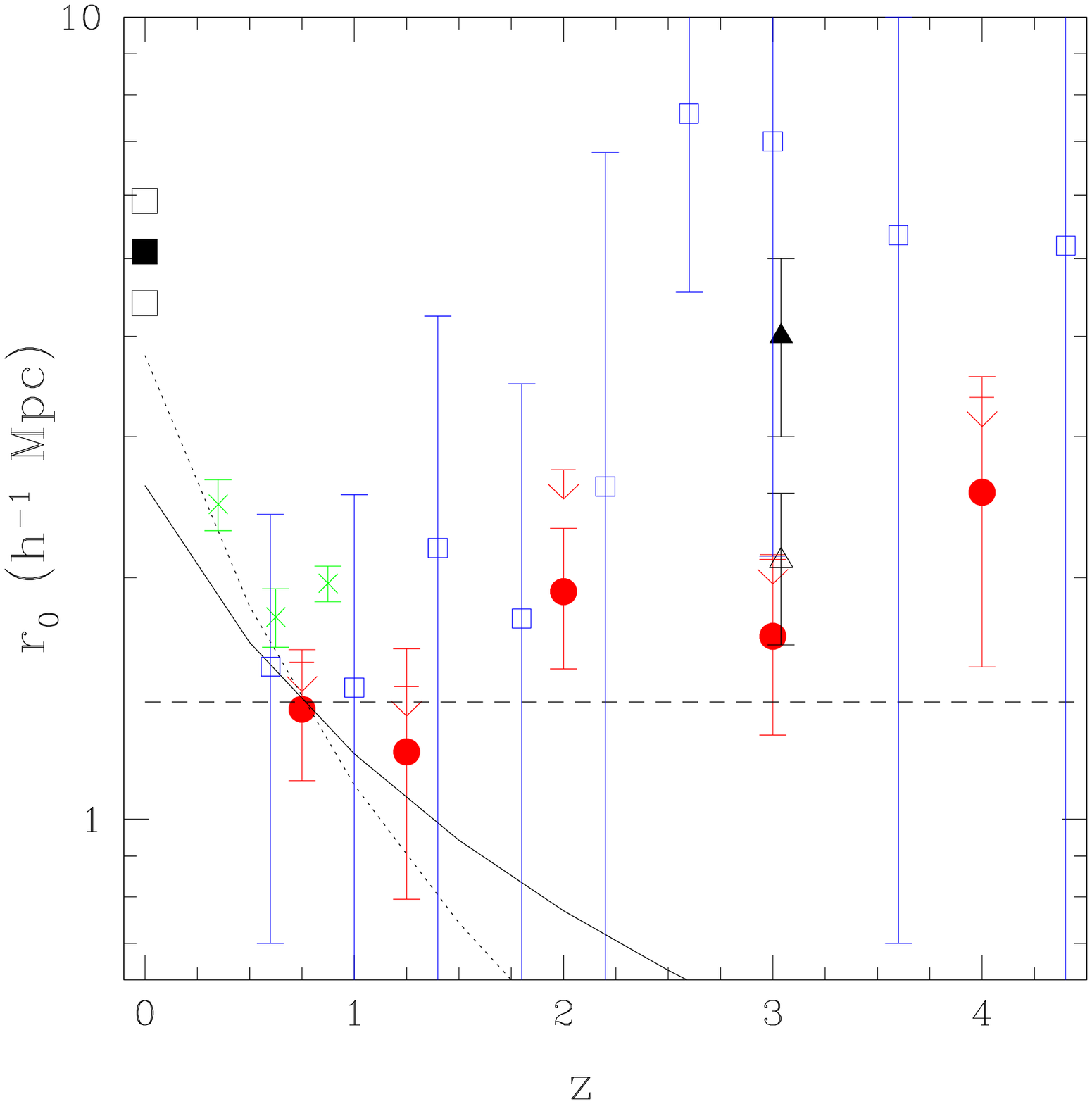,angle=0,width=7.5cm,height=7.5cm}
} 
\caption[]{
The measured comoving correlation length $r_0(z)$ (in $h^{-1}$ Mpc) as
obtained from the values of $A_{\omega}$ by assuming an EdS universe
(filled circles; the arrows refer to the upper-limits due to the
contamination). Previous determinations are shown by the same symbols
as in Figure~\ref{fawz}.  We add also the values obtained from the
analysis of the CFRS (Le F\`evre et al. 1996; cross symbols), from the
count-in-cell analysis of the LBG catalogue (Adelberger et al. 1998;
filled triangle) and the local values obtained from the APM survey
(Loveday et al.  1995; full sample, E/S0 and Sp/Irr sub-samples are
shown by the filled square, and the high and low open squares,
respectively). The curves show the evolution of the clustering using
the $\epsilon$ model with three values of the parameter $\epsilon$:
$\epsilon=0.8$ (linear growth of clustering, solid line); $\epsilon=2$
(growth more rapid than the linear prediction, dotted line),
$\epsilon=-1.2$ (fixed clustering in comoving coordinates, dashed
line).  The curves have been arbitrarily scaled to our observed value
at $z=0.75$. }
\label{frosigz} 
\end{figure}

The values for the comoving $r_0(z)$ obtained from our data are listed
in Table~\ref{tawz} for three different cosmologies.  In
Figure~\ref{frosigz}, we compare our values of $r_0$ as a function of
$z$ to a compilation of values taken from the literature.  The results
are given under the assumption of an EdS universe.  From this figure,
one can notice that $r_0$ shows a small decline from $z\simeq 0$ to
$z\simeq 1-1.5$ followed by an increase at higher $z$. At $z \ge 2$
the clustering amplitude is comparable to or higher than that observed
at $z\simeq 0.25$.

\begin{table*}
\centering
 \caption[]{The amplitude of $\omega(\theta)$ at 10 arcsec
 ($A_{\omega}$) for different redshift bins with $I_{AB}\le 28.5$.
 Column 1: the redshift range. 
 Column 2: the number of galaxies in the redshift bin. 
 Column 3: the amplitude of the ACF at 10 arcsec.
 Columns 4, 5, 6: the comoving correlation length $r_0$
 (in $h^{-1}$ Mpc) as derived from the Limber equation for three
 different cosmological models (EdS model, open model with
 $\Omega_{\rm 0m}=0.3$ and a flat model with
 $\Omega_{\rm 0m}=0.3$ and cosmological constant).
 All the listed values are not corrected for the contamination. 
             }
\begin{tabular}{ccccccc}
\hline
 $z$ range  & Number of & $A_{\omega}$ & $r_0$ & $r_0$ &
 $r_0$\\
  &  galaxies & (at 10 arcsec) &
 $\Omega_{\rm 0m}=1,\Omega_{\rm 0\Lambda}=0 $ &
 $\Omega_{\rm 0m}=0.3,\Omega_{\rm 0\Lambda}=0 $ & $
 \Omega_{\rm 0m}=0.3,\Omega_{\rm 0\Lambda}=0.7$ \\
\hline
0.0 - 0.5 & 96 & 0.17 $\pm$ 0.09 & 1.63 $\pm$ 0.47 & 1.77 $\pm$
0.51 & 1.93 $\pm$ 0.56 \\ 
0.5 - 1.0 & 294 & 0.09 $\pm$ 0.03 & 1.37 $\pm$ 0.25 & 1.69 $\pm$ 
0.31 & 1.93 $\pm$ 0.36 \\ 
1.0 - 1.5 & 157 & 0.09 $\pm$ 0.05 & 1.21 $\pm$ 0.41 & 1.64 $\pm$ 
0.56 & 1.82 $\pm$ 0.63 \\ 
1.5 - 2.5 & 202 & 0.12 $\pm$ 0.04 & 1.92 $\pm$ 0.38 & 3.06 $\pm$ 
0.61 & 3.07 $\pm$ 0.61 \\ 
2.5 - 3.5 & 142 & 0.13 $\pm$ 0.06 & 1.69 $\pm$ 0.41 & 3.06 $\pm$ 
0.75 & 2.78 $\pm$ 0.68 \\ 
3.5 - 4.5 & 35 & 0.35 $\pm$ 0.25 & 2.56 $\pm$ 1.01 & 5.29 $\pm$ 
2.08 & 4.28 $\pm$ 1.69 \\
0.0 - 6.0 & 959  &  0.03 $\pm$ 0.01 &         &        &   \\
\hline
\end{tabular}
\label{tawz}
\end{table*}

An implication of the results shown in this figure is that the
evolution of galaxy clustering cannot be properly described by the
standard parametric form: $\xi(r,z) = \xi(r,z=0) \
(1+z)^{-(3+\epsilon-\gamma)}$, where $\epsilon$ models the
gravitational evolution of the structures. Due to the dependence of
the bias on redshift and mass, the evolution of galaxy clustering is
related to the clustering of the mass in a complex way.  This has
already been noticed by G98 from the study of LBGs at $z\simeq 3$ (see
also Moscardini et al. 1998 for a theoretical discussion of the
problem).

In the plot of the correlation length $r_0$ we present also the
results for $z<1$ obtained by Le F\`evre et al. (1996) from the
estimates of the projected correlation function of the CFRS.  We do
not show in the figure the correlation lengths obtained by Carlberg et
al. (1997), who performed the same analysis using a K selected sample,
because they adopted a different cosmological model.  Their estimates
with $q_0=0.1$ of $r_0$ are approximately a factor of 1.5 larger than
the CFRS results in a comparable magnitude and redshift range.  Our
results are lower than these previous estimates and show that the
objects selected by our catalogue at low redshifts tend to have
different clustering properties. This effect suggests a dependence of
the clustering properties on the selection of the sample which is
even more evident at high redshift.  In fact our value of $r_0$ at $z
\simeq 3$ is smaller than that obtained by A98 and G98 for their LBG
catalogues at the same redshift. To measure the clustering properties,
A98 used a bright sample of 268 spectroscopically confirmed galaxies
and derived $r_0\simeq 4 \ h^{-1}$ Mpc; G98 used a larger sample of
871 galaxies and derived a value two times smaller ($r_0\simeq 2 \
h^{-1}$ Mpc). Our value, referring to galaxies with $I_{AB}\le 28.5$,
is $r_0\simeq 1.7\ h^{-1}$ Mpc.  Notice that this value is a lower
limit since it does not take into account the effects of
contamination.  All these reported values of the correlation length
are obtained by assuming an EdS universe.  This decrease of $r_0$
suggests that at fainter magnitudes we observe less massive galaxies
which are intrinsically less correlated. This is in qualitative
agreement with the prediction of the hierarchical galaxy formation
scenario (e.g.  Mo, Mao \& White 1999).  On the contrary, such an
interpretation is only marginally consistent with the reported higher
value at $z\simeq3$ of Magliocchetti \& Maddox (1999), computed with
the same FLY99 catalogue.

\begin{figure*}
\centerline{\psfig{figure=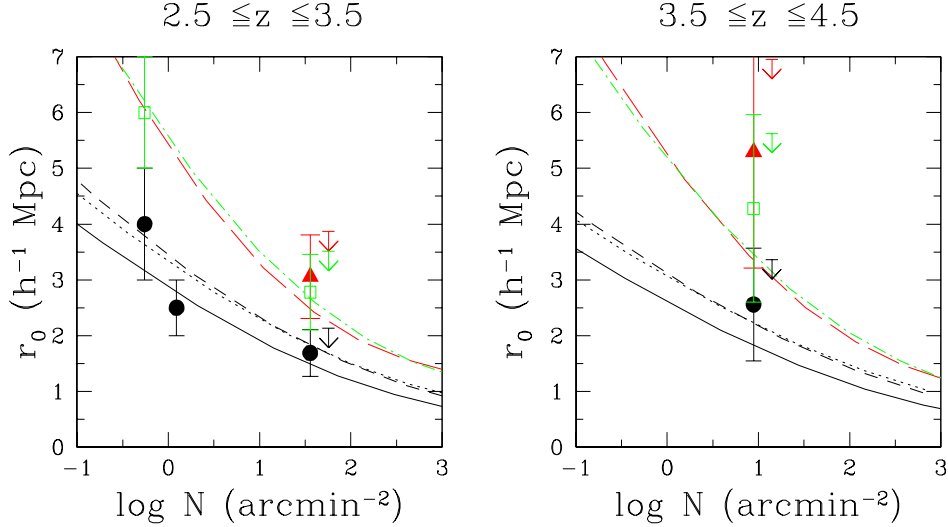,angle=0,width=13.cm,height=7.5cm}}
  \caption[]{The comoving correlation length $r_0$ (in $h^{-1}$ Mpc)
  as a function of surface density (defined as the expected number of
  objects per square arcminute) in two different redshift intervals:
  $2.5 \leq z \leq 3.5$ (left panel) and $3.5 \leq z \leq 4.5$ (right
  panel).  Different lines refer to the predictions of various
  cosmological models: SCDM (solid lines), $\tau$CDM model
  (short-dashed lines), TCDM (dotted lines), OCDM(long-dashed lines)
  and $\Lambda$CDM (dotted-dashed lines).  For comparison we show (at
  $\log N \simeq 1.55$ in the left panel and at $\log N \simeq 0.95$
  in the right panel) the results obtained in this work for three
  cosmologies: filled circle, filled triangle and open square refer to
  EdS models, open and flat universe with cosmological constant,
  respectively.  The arrows refer to the upper-limits due to
  contamination effects and are shifted of 0.2 in abscissa for
  clarity.  In the left panel we show also the results obtained from
  the analysis of the LBG clustering by Adelberger et al. (at $\log N
  \simeq -0.26$, for Einstein-de Sitter universe and for flat universe
  with cosmological constant), and from Giavalisco et al. (at $\log N
  \simeq 0.09$, only for EdS model).  }
\label{frodens} \end{figure*}

In order to better display the relation between the clustering
strength and the abundance of a given class of objects (defined as
haloes with mass larger than a given mass $M_{\rm min}$) in
Figure~\ref{frodens} we show, for the different cosmological models,
the relation between the predicted correlation length $r_0$ and the
expected surface density, i.e. the number of objects per square
arcminute. The quantity $r_0$ shown in this figure is defined as the
comoving separation where the predicted spatial correlation is unity;
the number density is computed by suitably integrating the modified
Press-Schechter formula [Equation (\ref{eq:ps2})] over the given
redshift range. In the left panel, showing the results for the
interval $2.5\leq z \leq 3.5$, we also plot the results obtained in
this work (points at high density with their associated upper-limits
due to contamination effects) and those coming from the LBG analysis
of A98 and G98 and corresponding to a lower abundance. All the models
are able to reproduce the observed scaling of the clustering length
with the abundance and no discrimination can be made between
them. Similar conclusions have been reached by Mo, Mao \& White
(1999).  The right panel shows the same plot but at $z\simeq 4$, where
the only observational estimates come from this work and from
Magliocchetti \& Maddox (1999).  Here the situation seems to be more
interesting. In fact the observed clustering is quite high and in the
framework of the hierarchical models seems to require a low abundance
for the relevant objects. This density starts to be in conflict with
the observed one (which represents a lower limit due to the unknown
effect of the selection function) for some of the models here
considered, for example the OCDM model.  Thus, if our results will be
confirmed by future observations, the combination of the clustering
strength and galaxy abundance at redshift $z\simeq 4$ could be a
discriminant test for the cosmological parameters.

\begin{figure*}
\centerline{
\psfig{figure=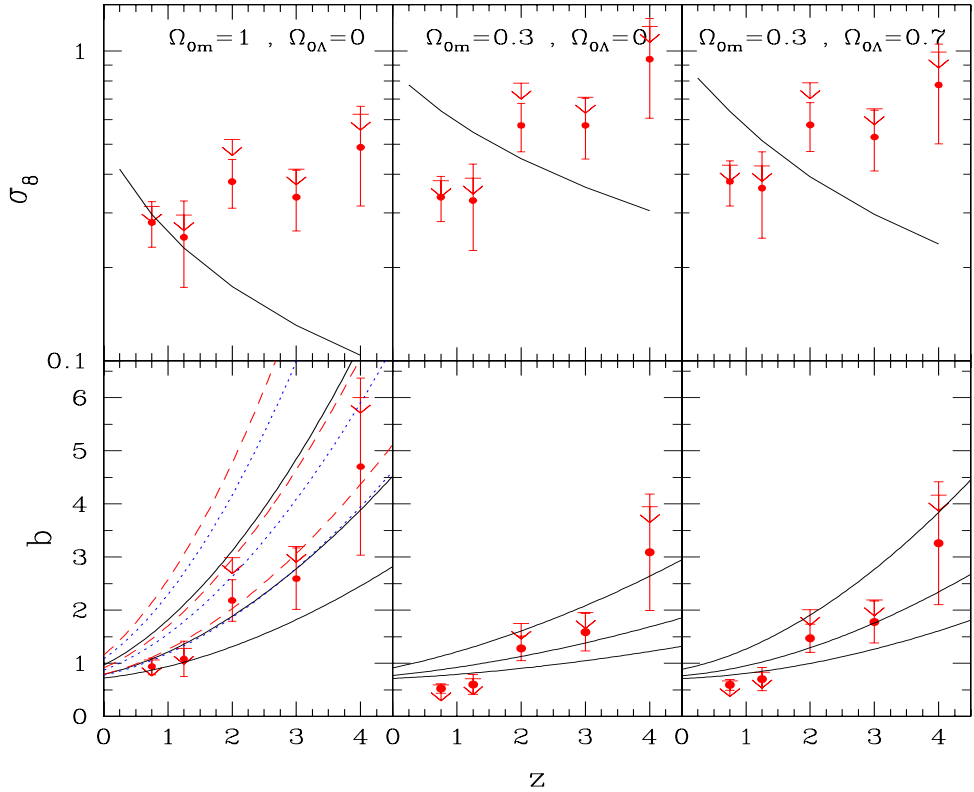,angle=0,width=14.cm,height=11.5cm}}
       \caption[]{The upper panels show (data as filled circles and
       upper-limits as arrows) our observed r.m.s. galaxy density
       fluctuation $\sigma_8^{\rm gal}(z)$ for three cosmologies
       (left: EdS universe; middle: open universe with $\Omega_{\rm
       0m}=0.3$ and vanishing cosmological constant; right: flat
       universe with $\Omega_{\rm 0m}=0.3$ and cosmological constant).
       The solid lines show the r.m.s. mass density fluctuation
       $\sigma_8^{\rm m}(z)$ for the same cosmologies, as obtained by
       linear theory.  The models are normalized to reproduce the
       local abundance of the galaxy clusters.  The lower panels show
       the measured bias $b$ as a function of redshift [with $b(z)
       \equiv \sigma_8^{\rm gal}(z)/\sigma_8^{\rm m}(z)$].  The curves
       show, for different cosmological models, the theoretical
       effective bias computed with different values of minimum mass.
       For the EdS universe three cosmological models are shown in the
       left panel: SCDM (solid lines), $\tau$CDM (dashed lines) and
       TCDM (dotted lines).  The central and right panels refer to
       OCDM and $\Lambda$CDM models, respectively.  We show results
       for $M_{\rm min}=10^{10}$, $10^{11}$ and $10^{12} h^{-1}
       M_{\odot}$ (lower curves are for smaller masses).  }
\label{fbias} \end{figure*}

An alternative way to study the clustering properties is given by the
observed r.m.s. galaxy density fluctuation $\sigma_8^{\rm gal}$. Its
redshift evolution is shown in the upper panels of Figure~\ref{fbias}
for three cosmological models: Einstein-de Sitter universe (left
panel); open universe with $\Omega_{\rm 0m}=0.3$ and vanishing
cosmological constant (central panel); flat universe with $\Omega_{\rm
0m}=0.3$ and cosmological constant (right panel). In the same plot we
show also the theoretical predictions computed by using the linear
theory when the cosmological models are normalized to reproduce the
local cluster abundance.  Since the corresponding values of
$\sigma_8^{\rm m}$ at $z=0$ (reported in Table \ref{t:models}) are
smaller than unity, we can safely compute the redshift evolution by
adopting linear theory.  As shown in Moscardini et al. (1998), the
differences between these estimates and those obtained by using the
fully non-linear method described above are always smaller than 3\% at
$z=0$ and consequently negligible at higher redshifts.  The comparison
suggests that, while some anti-bias is present at low redshift, the
high-redshift galaxies are strongly biased with respect to the dark
matter.  This observation strongly supports the theoretical
expectation of biased galaxy formation with a bias parameter evolving
with $z$.

Finally, the lower panels of Figure~\ref{fbias} report directly the
values of the bias parameter $b$ as deduced from our catalogue.  The
results show that $b$ is a strongly increasing function of redshift in
all cosmological models: from $z\simeq 0$ to $z\simeq 4$ the bias
changes from $b \simeq 1$ to $b \simeq 5$ in the EdS model and from $b
\simeq 0.5$ to $b \simeq 3$ in OCDM and $\Lambda$CDM models. This
qualitative behaviour is what is expected in the framework of the
hierarchical models of galaxy formation, as confirmed by the curves of
the effective bias computed by using Equation (\ref{eq:b_eff}), with
$M_{\rm min}=10^{10}$, $10^{11}$ and $10^{12} h^{-1} M_{\odot}$. The
observed bias is well reproduced when a minimum mass of $\simeq
10^{11} h^{-1} M_{\odot}$ is adopted for SCDM, in agreement with the
discussion of the results about the correlation amplitude
$A_\omega$. On the contrary, the study of the bias parameter for the
other two EdS models (TCDM and $\tau CDM$) seems to suggest a smaller
value of $M_{\rm min} \simeq 10^{10} h^{-1} M_{\odot}$. The
discrepancy is due to the fact that the computation of the correlation
amplitudes has been made by adopting a fixed slope of $\delta=0.8$,
which is not a good estimate of the best fit value for these two
models. For OCDM and $\Lambda$CDM models a minimum mass of $M_{\rm
min}\simeq 10^{11} h^{-1} M_{\odot}$ gives an effective bias in
agreement with the observations when $1.5 \mincir z \mincir 3$, while
a smaller (larger) minimum mass is required at lower (higher)
redshifts.

We can analyze the properties of the present-day descendants of our
galaxies at high $z$, assuming that the large majority of them
contains only one of our high-redshift galaxies (see e.g. Baugh et
al. 1998). Following Mo, Mao \& White (1999), we can obtain the
present bias factor of these descendants by evolving $b(z)$ backwards
in redshift from the formation redshift $z$ to $z=0$, according to the
`galaxy-conserving' model (Matarrese et al. 1997; Moscardini et
al. 1998); this gives
\be
b(M,0) = 1 + D_+(z)\bigl[b(M,z) - 1\bigr] \;, 
\ee
where, for $b(M,z)$, we can use the effective bias obtained for our
galaxies by dividing the observed galaxy r.m.s. fluctuation on
$8~h^{-1}$ Mpc by that of the mass, which depends on the background
cosmology.  For the galaxies at $z\simeq 3$ we find $b(M,0)\simeq 1.4,
1.3, 1.3$ for the EdS, OCDM and $\Lambda$CDM models, respectively.
The values of $b(M,0)$ that we obtained can be directly compared with
those for normal bright galaxies, which have $b_0\approx 1/\sigma_8$,
i.e. approximately 1.9 in the EdS universe and 1.1 in the OCDM and
$\Lambda$CDM models.  Consequently, the descendants of our galaxies at
$z\simeq 3$ appear in the EdS universe to be less clustered than the
present-day bright galaxies and can be found among field galaxies.  On
the contrary, the values resulting for the OCDM and $\Lambda$CDM
models seem to imply that the descendants are clustered at least as
much as the present-day bright galaxies, so they could be found among
the brightest galaxies or inside clusters. This is in agreement with
the findings of Mo, Mao \& White (1999) for the LBGs (see also Mo \&
Fukugita 1996; Governato et al. 1998; Baugh et al. 1999).  If we
repeat the analysis by using our galaxies at redshift $z\simeq 4$, we
find that $b(M,0)\simeq 1.8, 1.7, 1.6$ for the EdS, OCDM and
$\Lambda$CDM models, respectively. The ratio between the correlation
amplitudes of the descendants and the normal bright galaxies is
$\simeq 0.9, 2.3, 2.2$. This result confirms that for the EdS models
they have clustering properties comparable to ``normal'' galaxies,
while for non-EdS models the descendants seem to be very bright and
massive galaxies.

\section{Discussion and Conclusions}

In this paper we have measured over the redshift range $0 \le z \le
4.5$ the clustering properties of a faint galaxy sample in the HDF
North (Fern\'andez-Soto et al. 1999), by using photometric redshift
estimates. This technique makes it possible both to isolate galaxies
in relatively narrow redshift intervals, reducing the dilution of the
clustering signal (in comparison with magnitude limited samples;
Villumsen, Freudling \& da Costa 1997), and to measure the clustering
evolution over a very large redshift interval for galaxies fainter
than the spectroscopic limits.  The comparison with spectroscopic
measurements shows that, for galaxies brighter than $I_{AB}\le 26$,
our accuracy is close to $\sigma_z \sim 0.1$ for $z\le 1.5$ and
$\sigma_z \sim 0.2$ for $z\ge 1.5$.  We have checked the reliability
of our photometric redshifts in the critical interval $1.2 \le z \le
2$ by replacing the J, H, Ks photometry of Dickinson et al. (1999)
with the $F110W$, $F160W$ measurements in the HDF-N sub-area observed
with NICMOS (Thompson et al. 1999).  The new photometry is in general
consistent with the IR photometry of Fern\'andez-Soto et al. (1999)
and our photometric redshifts are not significantly changed.  In order
to infer the confidence level for the galaxies beyond the
spectroscopic limits ($26 \le I_{AB} \le 28.5$), we have compared our
results first with those obtained by other photometric codes and
second with Monte Carlo simulations.  The first comparison shows that
the resulting dispersion is $\sigma_z \simeq 0.20$ at $z\le 1.5$ and
increases at higher redshifts ($\sigma_z \simeq 0.30$), with a
possible systematic shift ($\langle z_{\rm GIS}-z_{\rm CE}\rangle
\simeq - 0.15$ and $\langle z_{\rm GIS}-z_{\rm FLY99}\rangle \simeq +
0.3$).  The second comparison with Monte Carlo simulations (made to
determine the effects of photometric errors in the redshift estimates)
shows that the r.m.s. dispersion obtained in this way is compatible
with the previous estimates done by comparing the different codes: for
galaxies with $I_{AB}\le 28.5$ we found $\sigma_z \simeq 0.2-0.3$ with
a maximum $\sigma_z=0.35$ for the redshift range $1.5\le z \le
2.5$. The contamination fraction of simulated galaxies incorrectly put
in a bin different from the original one due to photometric errors is
close to $\sim 20\%$.  The dominant source of contamination in a given
redshift bin is due to the r.m.s. dispersion in the redshift
estimates, with the exception of the bin $0\le z \le 0.5$ where the
contamination is due to the high-redshift galaxies ($z\ge 1$)
improperly put at low $z$. Due to the contamination effect at any
redshift, we note that our clustering measurements should be
considered as a lower limit.  Assuming that the contaminating
population is uncorrelated, we have applied a correction $(1-f)^2$ to
our original measurements, where $f$ is the contaminating
fraction. This correction should be regarded as an upper-limit.

As a consequence of the redshift uncertainties we have chosen to
compute the angular correlation function $\omega(\theta)$ in large
bins with $\Delta z = 0.5$ at $z\le 1.5$ and $\Delta z = 1.0$ at $z\ge
1.5$.  The resulting $\omega(\theta)$ has been fitted with a standard
power-law relation with fixed slope, $\delta=0.8$.  This latter value
can be questioned because of the present lack of knowledge about the
redshift evolution of the slope and its dependence on the different
classes of objects.  In order to avoid systematic biases in the
analysis of the results, the theoretical predictions have been treated
with the same basic assumptions.  The behaviour of the amplitude of
the angular correlation function at 10 arcsec ($A_{\omega}$) shows a
decrease up to $z\simeq 1-1.5$, followed by a slow increase.  The
comoving correlation length $r_0$ computed from the clustering
amplitudes shows a similar trend but its value depends on the
cosmological parameters.  Finally, we have compared our $\sigma^{\rm
gal}_8$ to that of the mass predicted for three cosmologies to
estimate the bias.  For all cases, we found that the bias is an
increasing function of redshift with $b(z\simeq 0)\simeq 1 $ and
$b(z\simeq 4) \simeq 5$ (for EdS universe), and $b(z\simeq 0) \simeq
0.5 $ and $b(z\simeq 4)\simeq 3 $ (for open and $\Lambda$ universe).
This result confirms and extends in redshift the results obtained by
Adelberger et al. (1998) and Giavalisco et al. (1998) for a
Lyman-Break galaxy catalogue at $z\simeq 3$, suggesting that these
high-redshift galaxies are located preferentially in the rarer and
denser peaks of the underlying matter density field.

We have compared our results with the theoretical predictions of a set
of different cosmological models belonging to the class of the CDM
scenario. With the exception of the SCDM model, all the other models
are consistent with both the local observations and the COBE
measurements. We model the bias by assuming that the galaxies are
associated in a one-to-one correspondence with their hosting dark
matter haloes defined by a minimum mass ($M_{\rm min}$). Moreover, we
assume that the haloes continuously merge into more massive ones.  The
values of $M_{\rm min}(z)$ used in these computations refer either to
a fixed mass or to the median mass derived by our GISSEL model or to
the value required to reproduce the observed density of galaxies at
any redshift.

The comparison shows that all galaxy formation models presented in
this work can reproduce the redshift evolution of the observed bias
and correlation strength.  The halo masses required to match the
observations depend on the adopted background cosmology. For the EdS
universe, the SCDM model reproduces the observed measurements if a
typical minimum mass of $ 10^{11} h^{-1}\ M_{\odot}$ is used, while
the $\tau$CDM and TCDM models require a lower typical mass of
$10^{10}-10^{10.5} h^{-1}\ M_{\odot}$. For OCDM and $\Lambda$CDM
models, the mass is a function of redshift, with $M_{\rm min} \le
10^{10} h^{-1}\ M_{\odot}$ at $z\le 1.5$, $10^{11.5} h^{-1}\
M_{\odot}$ between $1.5\le z\le 3$ and $10^{12} h^{-1}\ M_{\odot}$ at
$z\simeq 4$.  The higher masses required at high $z$ to reproduce the
clustering strength for these models are a consequence of the smaller
bias they predict at high redshifts compared to the EdS models.

We notice that at very low $z$, both OCDM and $\Lambda$CDM models
overpredict the clustering and consequently the bias. Two effects may
be responsible for this failure. First, the one-to-one correspondence
between haloes and galaxies may be an inappropriate description at low
$z$, where a more complex picture might be required. Second, we have
assumed that merging continues to be effective at low $z$ when, on the
contrary, the fast expansion of the universe acts against this
process, particularly for these models.

As a consequence of the bias dependence on the redshift and on the
selection criteria of the samples, the behaviour of the galaxy
clustering cannot provide a straightforward prediction on the
behaviour of the underlying matter clustering.  For this reason, the
parametric form $\xi(r,z) = \xi(r,z=0) (1+z)^{-(3+\epsilon-\gamma)}$,
where $\epsilon$ models the gravitational evolution of the structures,
cannot correctly describe the observations for any value of
$\epsilon$.

Another prediction of the hierarchical models is the dependence of the
clustering strength on the limiting magnitude of the samples.  At
$z\simeq 3$, we have compared our clustering measurements with the
previous results obtained for the LBGs by Adelberger et al. (1998) and
Giavalisco et al. (1998).  The three samples correspond to different
galaxy densities (our density in the HDF is approximately 65 times
higher than the LBGs of Adelberger et al. 1998).  The clustering
strength shows a decrease with the density.  This result is in
excellent agreement (both qualitatively and quantitatively) with the
clustering strength predicted by the hierarchical models as a function
of the halo density. More abundant haloes are less clustered than less
abundant ones (see also Mo et al. 1998). Moreover, this result, which
is independent of the adopted cosmology, supports our assumption of a
one-to-one correspondence between haloes and galaxies at
high-redshifts (see also Baugh et al. 1999), because otherwise we
would expect a higher small-scale clustering at the observed density.
As also noticed by Adelberger et al. (1998) for LBGs, such a result
seems to be incompatible with a model which assumes a stochastic star
formation process, which would predict that observable galaxies have a
wider range of masses. In fact, in this case the correlation strength
should be lower than the observed one because of the contribution by
the most abundant haloes (which are less clustered).  Moreover, it
seems possible to exclude that a very large fraction (more than 50\%)
of massive galaxies are lost by observations due to dust obscuration,
because the correlation strength would be incompatible with the
observed density.  Consequently, one of the main results of Adelberger
et al. (1998), namely the existence of a strong relation between the
halo mass and the absolute UV luminosity due to the fact that more
massive haloes host the brighter galaxies, seems to be supported by
the present work also for galaxies ten times fainter.

We have estimated the clustering properties at the present epoch of
the descendants of our high-redshift galaxies.  To do so, we have
assumed that only one galaxy is hosted by the descendants.  The
resulting local bias for the descendants of the galaxies at $z\simeq
3$ is $b(z=0)\simeq 1.4, 1.3, 1.3$ for the EdS, OCDM and $\Lambda$CDM
models, respectively.  Considering the galaxies at $z\simeq 4$, we
obtain $b(z=0)=1.8, 1.7, 1.6$, respectively.  These values seem to
indicate that in the case of the the EdS universe they are field or
normal bright galaxies while for the OCDM and $\Lambda$CDM models the
descendants can be found among the brightest and most massive galaxies
(preferentially inside clusters).  As already noted, at $z \simeq 3$,
the clustering strength and the observed density of galaxies are in
good agreement with the theoretical predictions for any fashionable
cosmological model.  At $z\simeq 4$, the present analysis seems to be
more discriminant.  Although our estimation should be regarded as
tentative and needs future confirmation, we find a remarkably high
correlation strength.  For some models the observed density of
galaxies starts to be inconsistent with the required theoretical halo
density.  The relation between clustering properties and number
density of very high redshift galaxies therefore provides an
interesting way to investigate the cosmological parameters.  The
difference in the predicted masses ($\simeq$ 15 to 30 at $z\simeq$ 3
and 4) between EdS and non-EdS universe models is also in principle
testable in terms of measured velocity dispersions.  The present
results have been obtained in a relatively small field for which the
effects of cosmic variance may be important (see Steidel 1998 for a
discussion).  Nevertheless they show a possibility of challenging
cosmological parameters which becomes particularly exciting in view of
the rapidly growing wealth of multi-wavelength photometric databases
in various deep fields and availability of 10m-class telescopes for
spectroscopic follow-up in the optical and near infrared.
  
\section*{Acknowledgments.} 
We are grateful to H. Aussel, C. Benoist, M. Bolzonella, A. Bressan,
S. Charlot, S. Colombi, S. d'Odorico, H. Mo, R. Sheth and G. Tormen
for useful general discussions.  We acknowledge K. Lanzetta,
A. Fern\'andez-Soto and A. Yahil for having made available the
photometric optical and IR catalogues of the HDF North.  We also thank
the anonymous referee for comments which allowed us to improve the
presentation of this paper.  Many thanks to P. Bristow for carefully
reading the manuscript.  This work was partially supported by Italian
MURST, CNR and ASI and by the TMR programme Formation and Evolution of
Galaxies set up by the European Community.  S. Arnouts has been
supported during this work by a Marie Curie Grant Fellowship.

\end{document}